\documentclass[aps,prd,twocolumn,showpacs,superscriptaddress,groupedaddress,ref,floatfix]{revtex4}  
\usepackage{graphicx}  
\usepackage{dcolumn}   
\usepackage{bm}        
\usepackage{amssymb}   
\usepackage{amsmath} 
\usepackage{natbib} 
\usepackage{aas_macros} 
\usepackage{here}  
\pdfoutput=1
\begin{document}

% The following information is for internal review, please remove them for submission
\widetext
%\leftline{Version 01 as of \today}
%\leftline{Primary author: Milad Delfan Azari}
%\leftline{To be submitted to (PRD)}
%\leftline{Comment to {\tt d0-run2eb-nnn@fnal.gov} by xxx, yyy}
%\centerline{\em D\O\ INTERNAL DOCUMENT -- NOT FOR PUBLIC DISTRIBUTION}

% the following line is for submission, including submission to the arXiv!!
%\hspace{5.2in} \mbox{Fermilab-Pub-04/xxx-E}

\title{Linear Analysis of Fast-Pairwise Collective Neutrino Oscillations in Core-Collapse Supernovae based on the Results of Boltzmann Simulations}

\author{Milad Delfan Azari}

\affiliation{Department of Physics, Graduate School of Advanced Science and Engineering, Waseda University, 3-4-1 Okubo, Shinjuku, Tokyo 169-8555, Japan}

\author{Shoichi Yamada}

\affiliation{Department of Physics, Graduate School of Advanced Science and Engineering, Waseda University, 3-4-1 Okubo, Shinjuku, Tokyo 169-8555, Japan}

\affiliation{Advanced Research Institute for Science and Engineering, Waseda University, 3-4-1 Okubo, Shinjuku, Tokyo 169-8555, Japan}

\author{Taiki Morinaga}

\affiliation{Department of Physics, Graduate School of Advanced Science and Engineering, Waseda University, 3-4-1 Okubo, Shinjuku, Tokyo 169-8555, Japan}

\author {\mbox{Wakana Iwakami}} 

\affiliation{Advanced Research Institute for Science and Engineering, Waseda University, 3-4-1 Okubo, Shinjuku, Tokyo 169-8555, Japan}

\affiliation{Yukawa Institute for Theoretical Physics, Kyoto University,Oiwake-cho, Kitashirakawa, Sakyo-Ku, Kyoto, 606-8502, Japan}

\author {Hirotada Okawa} 

\affiliation{Advanced Research Institute for Science and Engineering, Waseda University, 3-4-1 Okubo, Shinjuku, Tokyo 169-8555, Japan}

\affiliation{Yukawa Institute for Theoretical Physics, Kyoto University,Oiwake-cho, Kitashirakawa, Sakyo-Ku, Kyoto, 606-8502, Japan}

\affiliation{Waseda Institute for Advanced Study, 1-6-1 Nishi Waseda, Shinjuku, Tokyo 169-8050, Japan}

\author {Hiroki Nagakura} 

\affiliation{Department of Astrophysical Sciences, Princeton University, Princeton, NJ 08544, USA}

\author {Kohsuke Sumiyoshi} 

\affiliation{Numazu College of Technology, Ooka 3600, Numazu, Shizuoka 410-8501, Japan}

\date{\today}

\begin{abstract}

Neutrinos are densely populated deep inside the core of massive stars after their gravitational \mbox{collapse} to produce supernova explosions and form compact stars such as neutron stars (NS) and black holes (BH). It has been considered that they may change their flavor identities through \mbox{so-called} fast-pairwise conversions induced by mutual forward scatterings. If that is really the case, the dynamics of supernova explosion will be influenced, since the conversion may occur near the neutrino sphere, from which neutrinos are effectively emitted. In this paper, we conduct a \mbox{pilot} study of such possibilities based on the results of fully self-consistent, realistic simulations of a core-collapse supernova explosion in two spatial dimensions under axisymmetry. As we solved the Boltzmann equations for neutrino transfer in the simulation not as a post-process but in real time, the angular distributions of neutrinos in momentum space for all points in the core at all times are available, a distinct feature of our simulations. We employ some of these distributions extracted at a few selected points and times from the numerical data and apply linear analysis to assess the possibility of the conversion. We focus on the vicinity of the neutrino sphere, where different species of neutrinos move in different directions and have different angular distributions as a result. This is a pilot study for a more thorough survey that will follow soon. We find no positive sign of conversion unfortunately at least for the spatial points and times we studied in this particular model. We hence investigate rather in detail the condition for the conversion by modifying the neutrino distributions rather arbitrarily by hand. 
     
\end{abstract}

%\pacs{}
\maketitle{}

\section{introduction}

Neutrinos are massive particles although their masses are much smaller than those of the charged lepton counterparts \cite{1998PhRvL..81.1562F}. Moreover, the masses are not diagonal with respect to the flavors and, as a consequence, the flavor conversion, or the neutrino oscillation, occurs as they propagate in vacuum \cite{2008PhRvD..78h3007G}. In the presence of matter, weak interactions with surrounding matter modify this dispersion relation in vacuum and induce resonant conversions of flavors, which are refered to as the MSW (Mikheyev-Smirnov-Wolfenstein) effect and believed to be the solution to the solar neutrino problem \cite{Mikheev:1987qk,1978PhRvD..17.2369W}. The neutrino self-energy induced by weak interactions, which is the origin of the MSW effect, is also generated by the interactions with other neutrinos and if they are densely populated, it will contribute to the flavor conversion and is called the collective neutrino oscillation \cite{1992PhLB..287..128P,2007PhRvD..75h3002R,2008PhRvD..78h5012E,2010ARNPS..60..569D}. This is qualitatively different from the former two, since the equations that describe the propagation of neutrinos become nonlinear and, as a result, interesting new phenomena such as spectral splittings may occur \cite{2007PhRvD..76h1301R,2015IJMPE..2441008D,2016NuPhB.908..366C}.

By definition the collective neutrino oscillation occurs only where neutrinos are abundant. Core-collapse supernovae (CCSNe) are one of such sites in the universe as vindicated by the detection of about twenty electron-type anti-neutrinos from the supernova SN1987A in the Large Magellanic Cloud \cite{1987PhRvL..58.1490H}. CCSNe are the explosive death of massive stars with zero-age-main-sequance (ZAMS) masses of $\gtrsim 8M_{\odot}$ and are at the same time the birth of a compact object such as neutron star (NS) or black hole (BH). They are also an important agent for the chemical evolution of the universe, producing heavy elements. The exact mechanism of CCSNe is not fully understood, though \cite{2012ARNPS..62..407J}. The initial implosion of a massive star core should be reversed somehow to produce an explosion. It is well established that the core bounce induced by hardening of matter at the nuclear density does not generate a shock wave powerful enough to expell the outer part of the imploding core, not to mention the stellar outer envelopes. Supernova researchers have been hence seeking for a way to reinvigorate the shock wave stalled inside the core.

Neutrinos ($\nu$'s) are believed to play a key role in the shock revival. As a matter of fact, almost all of the binding energy of NS liberated in the gravitational collapse is emitted in the form of neutrinos and the kinetic energy of matter in the supernova explosion is just $\sim{1\%}$ of this energy. In the currently most popular scenario, which is called the neutrino heating mechanism, a fraction of the electron-type neutrinos and anti-neutrinos are re-absorbed by the matter between the shock front and the so-called gain radius and deposit their energy to push the stagnated shock again. It is obvious then that the success of the scenario depends on how efficiently neutrinos are absorbed by matter. It is also known that $\nu_{\mu}$, $\nu_{\tau}$ and their anti-particles have higher energies than $\nu_{e}$ and $\bar{\nu}_e$ in general, since they lack the interactions with matter via charged currents in the supernova core. It is equally obvious then that if the former is converted to the latter and absorbed by matter, more energy will be transferred to matter and may induce successful explosions. This is one of the reasons why the collective neutrino oscillation is attracting the interest of supernova researchers and particle physicists alike. In fact the interest is revived when it is realized that the so-called fast-pairwise conversion may occur near the neutrino sphere, which has a radius that roughly corresponds to the optical depth of 1 and is the surface, from which neutrinos are effectively emitted \cite{2016PhRvL.116h1101S,2009PhRvD..79j5003S}.

The investigation of the collective neutrino oscillations is much more difficult than those of the vacuum or MSW oscillations, since the former is nonlinear phenomena as already mentioned. As a result, the previous studies of the fully nonlinear oscillations were limited to some simplified and/or idealized situations \cite{2013PhRvD..88l5008D,2010ARNPS..60..569D,2006PhRvD..74j5014D,Hannestad:2006nj,2010PhRvD..81g3004D,Raffelt:2013isa,Abbar:2018beu}. The investigations of more realistic settings such as those in the supernova core are even more difficult, since kinetic equations that describe the neutrino transfer in non-uniform matter should be solved somehow. This is not an easy task even in spherical symmetry\cite{2009JPhG...36k3201D,2012PhRvD..85k3007S,2015PhRvD..91b5001P}.

Recently, a different approach based on linear analysis has been employed \cite{2012PhRvD..86h5010M,2011PhRvD..84e3013B,2012PhRvD..85k3002S,2012PhRvL.108f1101S}. The idea is based on the fact that neutrinos are almost in the flavor eigenstates at the beginning of the conversion; then the linearized equations can be used to study where and when the flavor conversion is triggered. In this approach the flavor conversion is regarded as the instability of the flavor eigenstate. It is a common practice to assume also the local approximation, in which the distributions of the background matter and neutrinos are uniform in space. This is of course justified only when the oscillation length is much shorter than the local scaleheights in the matter and neutrino distributions. The linearization of the original nonlinear equations makes the analysis drastically easier although we need to carefully handle spurious modes that plague numerical solutions of the linearized equations more often than not \cite{2018PhRvD..97b3024M}. More recently, it is demonstrated that the so-called dispersion-relation approach is more convenient \cite{2017PhRvL.118b1101I}. In this paper we base our analysis on this method.  

Quantitative studies of the fast-pairwise conversions in the realistic settings have been mostly limited to spherically symmetric 1D models so far \cite{2005PhRvD..72d5003S,2009PhRvD..79j5003S,2016PhRvL.116h1101S,2017JCAP...02..019D,2018arXiv180703322D}. Unfortunately they found no positive result \cite{2017JCAP...02..019D,2018PhRvD..97b3017D} except for the lowest-mass end of massive stars, which are supposed to produce the so-called electron-capture core-collapse supernovae \cite{2018PhRvD..98d3014A}. Very recently, Abbar et al.~\cite{Abbar:2018shq} extended such studies to 2D and 3D models. They extracted three snapshots from numerical data and looked for the crossing in the angular distributions of $\nu_{e}$ and $\bar\nu_{e}$. They found a positive sign in extended regions with the radius of $\gtrsim~50-70$~km. They also conducted a linear analysis, assuming that the angular distributions are axisymmetric with respective to the local radial direction and estimated the growth rate in the same direction. Note, however, that these neutrino distributions were obtained by the neutrino transport calculation done as a post process, dropping the time dependence of both hydrodynamical and neutrino quantities and ignoring matter motion entirely. Hence they are not fully self-consistent.

In this paper we conduct a linear stability analysis for some selected neutrino distributions obtained in fully self-consistent simulations of CCSNe in two spatial dimensions under axisymmetry with our Boltzmann-radiation-hydrodynamics code~\cite{2018ApJ...854..136N}. Computing neutrino transport in situ together with hydrodynamics, they are five dimensional in fact (2 for space and 3 for momentum space) and are hence fully consistent with matter dynamics. This paper is actually meant to be a pilot study for more thorough investigations of the possibility of fast-pairwise conversion in our realistic models (\mbox{Morinaga et al. in preparation}). 

We pay particular attention to the vicinity of the neutrino sphere (r $\lesssim50$km) in this work. This is mainly because non-spherical features are most manifest there: neutrino distributions are not axisymmetric with respect to the radial direction unlike in 1D; the neutrino fluxes are non-radial and not aligned with each other among different species. Note that these asymmetries are mainly produced by convective matter motions near the neutrino sphere; the neglect of matter motion may hence understimate them. Another reason is, of course, that if the flavor conversion occurs near the neutrino sphere, it will have the largest impact on the supernova dynamics as originally emphasized by Sawyer~\cite{2016PhRvL.116h1101S}. Note that in this paper we will not just look for the crossing in the angular distributions of $\nu_{e}$ and $\bar\nu_{e}$, since it remains to be demonstrated that the crossing is really the condition for the fast-pairwise conversion particularly in multi-D settings (see \cite{2017PhRvL.118b1101I, 2017JCAP...02..019D, 2018PhRvD..97b3017D, 2017PhRvD..96d3016C}). We hence conduct linear analysis for highly non-radial angular distributions irrespective of the exitence of crossing. As a matter of fact, when the original distributions fail to give conversion, we modify them until we find conversion and study the condition for successful conversion. 

This article is organized as follows. In \mbox{Section II}, we briefly review the EOM for neutrino flavors and formulate the linear stability analysis based on the dispersion relation. In Section III, we introduce the numerical data extracted from our realistic simulation of CCSN and employed for the linear analysis as the background models in this study. In Section IV, we present the results from the pilot study on the possibility of the fast-pairwise conversion and discuss in detail the condition for the instability. Finally in Section V, we summarize our results and conclude the paper.

\section{Formulation of Linear Stability Analysis}

\subsection{Equations of Motion (EOM)}

Following the previous works \cite{2017PhRvL.118b1101I,2018PhRvD..97b3024M}, we write down the EOM for the {$n_F\times n_F$} density matrix  $\rho$, which describes flavor evolutions of the neutrinos that have \mbox{energy $E$} and propagate in a particular direction. Here \mbox{$n_F$ is} the number of neutrino flavors. The diagonal components of the density matrix are the distribution functions of those neutrinos in the individual flavor eigenstates. The off-diagonal elements, on the other hand,  express the phase information in the oscillation from one flavor to another.
If ordinary collisional processes are neglected, the EOM can be written as 
 \begin{equation}
(\partial_{t}+\mathbf{v}\cdot\boldsymbol{\nabla}_{\boldsymbol{r}})\rho=i[\rho,H],\label{freestreaming}
\end{equation}
in which the Hamiltonian is expressed as

\begin{equation}
H=\dfrac{M^{2}}{2E}+v^{\mu}\Lambda_{\mu}\dfrac{\sigma_{3}}{2}+\sqrt{2}G_{F}\int d\Gamma'v^{\mu}v_{\mu}'\rho',\label{Ham}
\end{equation}
where $M^2$ is the mass-squared matrix, which causes the flavor oscillations in vacuum, $\sigma_3$ is one of the Pauli matrices and $v^{\mu}=(1,\mathbf{v})$ denotes the neutrino four velocity; the matter potential, which induces the MSW oscillation, is given as a four vector $\Lambda^{\mu}=\sqrt{2}G_{F}(n_{e}-n_{e^+})$ $u^{\mu}$, since matter is moving at the four velocity of $u^{\mu}$; note that we will work in the laboratory frame in the following; the last term is responsible for the collective oscillation and $\rho'$ is the density matrix for the neutrinos having energy $E'$ and moving at the four velocity of $v'^{\mu}$ = $(1, \mathbf{v'})$ and the integration is done over the whole momentum space $d{\Gamma'} = d\mathbf{v'}/4\pi$.

In the rest of the paper, we work in the two-flavor ($\nu_{e}$ and $\nu_{x}$) approximation as a common practice for simplicity \cite{Abbar:2018shq}, where $\nu_{x}$ stands for $\nu_{\mu}$ and $\nu_{\tau}$ collectively. Then we express the density matrix as  
\begin {equation}
\rho=\dfrac{f_{\nu_{e}}+f_{\nu_{x}}}{2}+\dfrac{f_{\nu_{e}}-f_{\nu_{x}}}{2}\begin{pmatrix}s & S\\
S^{*} & -s
\end{pmatrix},
\end{equation}
where $f_{\nu_{e}}$ and $f_{\nu_{x}}$ are the neutrino distribution functions for $\nu_{e}$ and $\nu_{x}$, respectively. Note that $s$ and $S$ are $\pm{1}$ and $0$, respectively, when the neutrino is in one of the flavor eigenstates. This simple fact suggests us to employ linear analysis. In fact, since neutrinos are produced in one of the flavor eigenstates, the off-diagonal components should be much smaller than the diagonal ones until the flavor conversion is triggered and the former components grow exponentially. Then the use of the linearized EOM for the small off-diagonal components will be justified in the study on the trigger of the flavor conversion. 
\subsection{Linear Stability Analysis Based On Dispersion Relation}

Note first that the flavor eigenstates corresponding to $s=\pm{1}$ and $S=0$ are fixed points of EOM if we ignore the off-diagonal elements in the mass matrix $M$ in vacuum, which we will assume in the following indeed. Then by linearizing Eq. (\ref{freestreaming}) at one of these fixed points, we obtain the following EOM for the small off-diagonal component $S_\mathbf{v}(E)$: 
\begin{equation}
i(\partial_{t}+\mathbf{v}\cdot\boldsymbol{\nabla}_{\mathbf{r}})S_{\mathbf{v}}=v^{\mu}(\Lambda_{\mu}+\Phi_{\mu})S_{\mathbf{v}}-\int\frac{d\mathbf{v}'}{4\pi}v^{\mu}v_{\mu}'G_{\mathbf{v'}}S_{\mathbf{v'}}\label{EQNLIN},
\end{equation}
where we add the subscript ${\bold v}$ to $S$ to indicate explicitly that it depends on $\mathbf v$; $G_{\mathbf{v}}$ is the electron-lepton number (ELN) angular distribution defined as 
\begin{equation}
G_{\mathbf{v}}=\sqrt{2}G_{F}\int_{0}^{\infty}\frac{dEE^{2}}{2\pi^{2}}\left[f_{\nu_{e}}(E,\mathbf{v})-f_{\bar{\nu}_{e}}(E,\mathbf{v})\right].\label{gv}
\end{equation}
The corresponding ELN current $\Phi^{\mu}$ can be defined as  
\begin{equation}
\Phi^{\mu}\equiv\int\frac{d\mathbf{v}}{4\pi}G_{\mathbf{v}}v^{\mu}.
\end{equation}

By assuming the solution of Eq.~(\ref{EQNLIN}) in the form of \mbox{$S_{\mathbf{v}}=Q_{\mathbf{v}}e^{-i(\Omega t-\mathbf{K}\cdot\mathbf{r})}$}, the equation for the amplitude $Q_{\mathbf{v}}$ is written as 
\begin{equation}
v^{\mu}k_{\mu}Q_{\mathbf{v}}=-\int\frac{d\mathbf{v}'}{4\pi}v^{\mu}v_{\mu}'G_{\mathbf{v}'}Q_{\mathbf{v}'},\label{EigenEq2}
\end{equation}
where $k^{\mu}=K^{\mu}-\Lambda^{\mu}-\Phi^{\mu}$ with $k^{\mu}=(\omega,\mathbf{k})$ and \mbox{$K^{\mu}=(\Omega,\mathbf{K})$}. Since the right-hand side of Eq. (\ref{EigenEq2}) can be expressed as $v^{\mu}a_{\mu}$ with
\begin{equation}
a^{\mu}\equiv-\int\frac{d\mathbf{v}}{4\pi}v^{\mu}G_{\mathbf{v}}Q_{\mathbf{v}},
\end{equation}
we can write $Q_{\mathbf{v}}=v^{\mu}a_{\mu}/v^{\mu}k_{\mu}$. If we put this expression back into Eq. (\ref{EigenEq2}), we obtain the equation
\begin{equation}
\Pi^{\mu\nu}(\omega,\mathbf{k})a_{\nu}=0\label{PI},
\end{equation}
where $\Pi^{\mu\nu}$ is given as 
\begin{eqnarray}
\Pi^{\rho\sigma} & =&\eta^{\rho\sigma}+\int\frac{d\mathbf{v}}{4\pi}G_{\mathbf{v}}\frac{v^{\rho}v^{\sigma}}{\omega-\mathbf{v}\cdot\mathbf{k}}\nonumber. \\
\end{eqnarray} 
Equation (\ref{PI}) has non-trivial solutions if and only if 
\begin{equation}
D(\omega,\mathbf{k})\equiv\det\Pi=0.\label{DR}
\end{equation} 
This gives implicitly a relation between $\omega$ and $\mathbf{k}$ that is referred to as the dispersion relation. Note that it depends on the direction of $\mathbf{k}$ in general although $\mathbf k$ is often assumed to be radial in the literature. Our numerical code can treat an arbitrary direction of $\mathbf k$ and we show later that not the radial direction but the so-called crossing direction is more important indeed when the fluxes of $\nu_e$ and $\bar\nu_e$ are misaligned with the radial direction.

We are interested in complex solutions of Eq.~(\ref{DR}), since they are supposed to indicate the instability corresponding to the initiation of the flavor conversion. Some cautions are necessary, however. As pointed out by \cite{Lingwood} and actually has been known for decades in plasma physics \cite{1964JAP....35.3268B,1981phki.book.....L}, it is not sufficient to obtain complex solutions with imaginary parts of appropriate signs in $\omega$ or $k$. In fact, the mathematically rigorous criterion for instability is the coalescence of two roots that are originated in the opposite halves of the complex \mbox{$k$-plane} in some moving frame, which is not easy to apply to realistic dispersion relations. It should be also noted that the coalescence search must be done recursively if the spatial dimension is larger than 1, which is practically impossible. In this paper we will be hence satisfied with the following: we regard complex $\omega$ solutions with positive imaginary parts for real $k$ as the sign of the flavor conversion; in some cases we also study the motions of some solutions in the complex $k$-plane as we change the value of the imaginary part of $\omega$. This is certainly an approximation and should be improved one way or another in the future (Morinaga et al. in preparation).  

\section{application to realistic data}

\subsection{Numerical Models}

We use the results of our realistic two-dimensional (2D) simulations on the K supercomputer systems in Japan \cite{2018ApJ...854..136N}. The core-collapse and the subsequent time evolution were computed fully self-consistently for the non-rotating progenitor model of 11.2$M_{\odot}$ in \cite{2002RvMP...74.1015W} with the Boltzmann equations for neutrino transport being solved by applying the discrete-oridinate method and taking fully into account special relativistic effects with a two-energy grid technique \cite{2014ApJS..214...16N}. In addition the Newtonian hydrodynamical equations and the Poisson equation for self-gravity were solved simultaneously. All the equations are written on spherical coordinates $(r,\theta)$ under the assumption of spatial axisymmetry. The computational domain covers the region of $0\leq r\leq5000$ km and $0\leq\theta\leq\pi$ in space, which is divided into 384$(r)\times$ 128$(\theta)$ mesh cells; momentum space was also discretized with 20 energy bins over the range from 0 to 300 MeV and with 10($\theta_{\nu}$)$\times$6($\phi_{\nu}$) angular mesh cells on the entire solid angle. Three neutrino species: electron-type neutrino $\nu_{e}$, electron-type anti-neutrino $\bar\nu_{e}$ and all the others $\nu_x$ were considered.

Although we employed two realistic equations of state (EOS) in the original simulations for comparison, we adopt in this paper only the results obtained for Furusawa's equation of state (FSEOS) \cite{2011ApJ...738..178F,2013ApJ...772...95F}, in which greater misalignments tend to be produced among the angular distributions of different neutrino species in the early post-bounce phase than in the other \mbox{equation of state by Lattimer \& Swesty (LSEOS) \cite{1991NuPhA.535..331L}.}

\begin{figure}[H]
\begin{tabular}{c} 
\includegraphics[width=6.2cm]{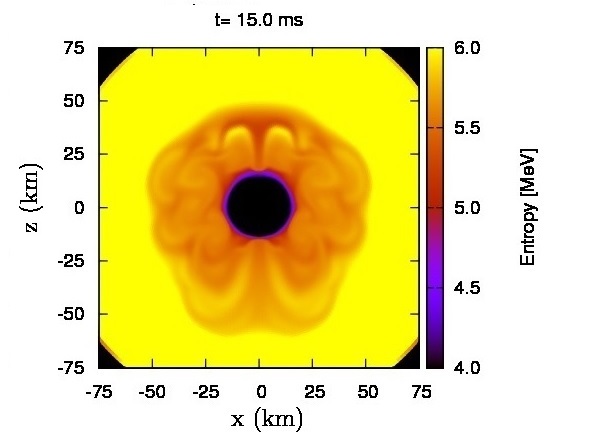} \\
\includegraphics[width=6.2cm]{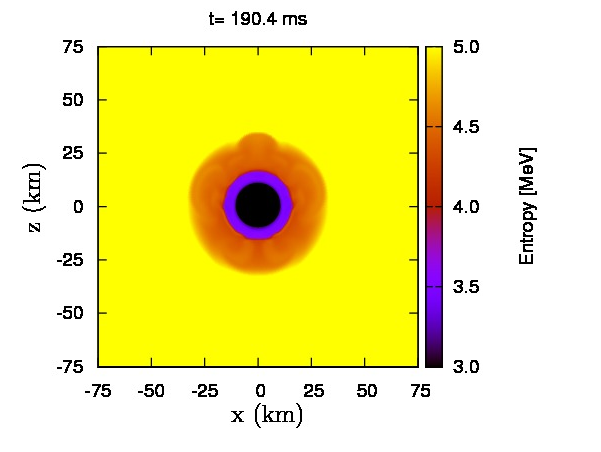}\\
\includegraphics[width=6.2cm]{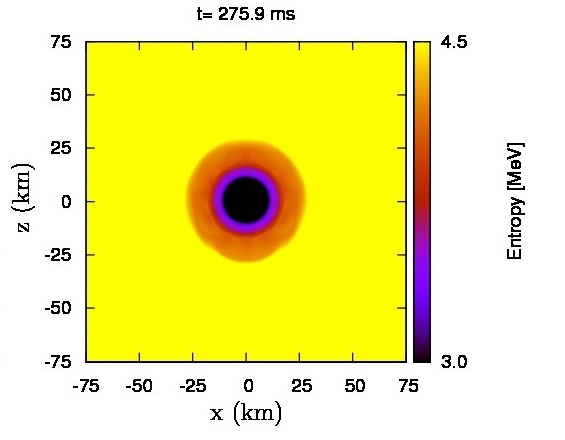}\\ 
\end{tabular} \caption{\label{ent15} The entropy distributions in the meridian section of the central part of the core at $t_{pb}$ = 15.0 ms (top), \mbox{$t_{pb}$ = 190.4 ms (middle)} and $t_{pb}$ = 275.9 ms (bottom), \mbox{respectively.}}
\end{figure} 

 We then picked up three snapshots at different times: $t_{pb}$ = 15.0, 190.4 and 275.9 ms post-bounce. \mbox{Figure~\ref{ent15}} displays the entropy distributions in the meridian section of the central part of the core at these times. The strong prompt convection is clearly seen in the top panel for $t_{pb}$=~15.0 ms although the shock front, which is visible near the corners, is still almost spherical. The shock wave is stagnated then and the so-called \mbox{neutrino-driven convection} sets in thereafter in the gain region, in which net neutrino heating occurs, whereas the convection near the PNS is subsided. The middle panel is the consequence of these evolutions up to \mbox{$t_{pb}$ = 190.4 ms}. At the even later time of \mbox{$t_{pb}$ = 275.9 ms} (bottom panel), both the convection near the PNS and the shock instability calm down and the shock front recedes to a smaller radius. This model is unlikely to produce an explosion.

\begin{figure}[t]
\begin{tabular}{c} 
\includegraphics[width=6.1cm]{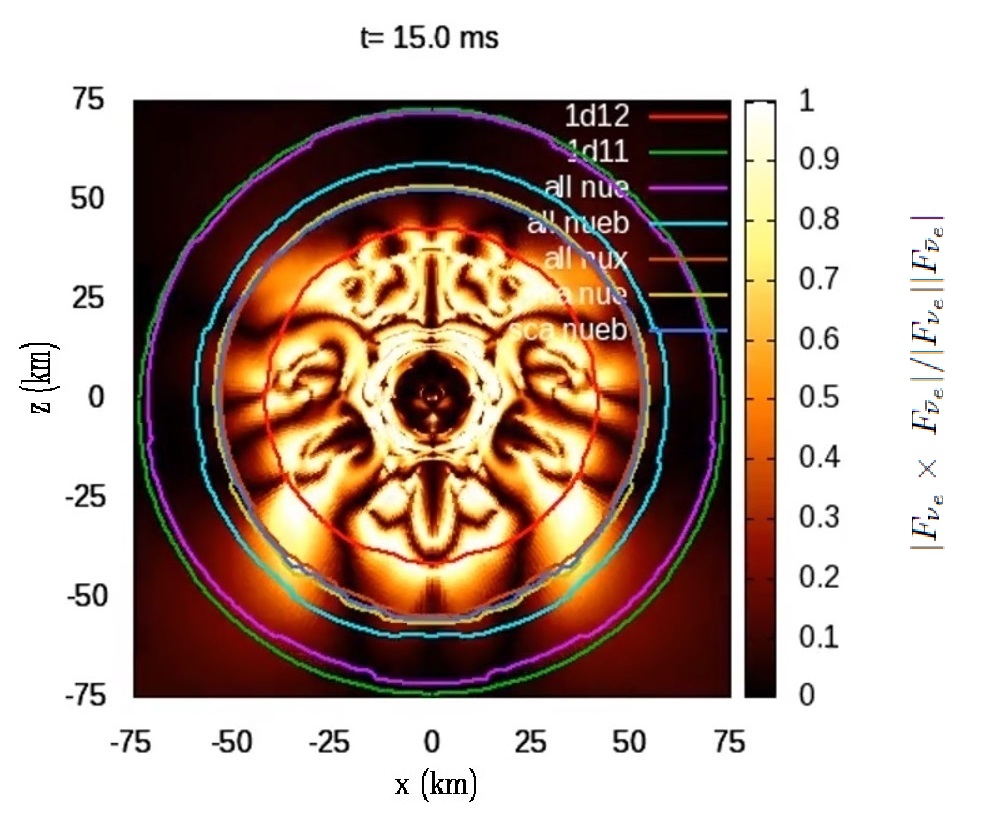} \\ 
\includegraphics[width=6.1cm]{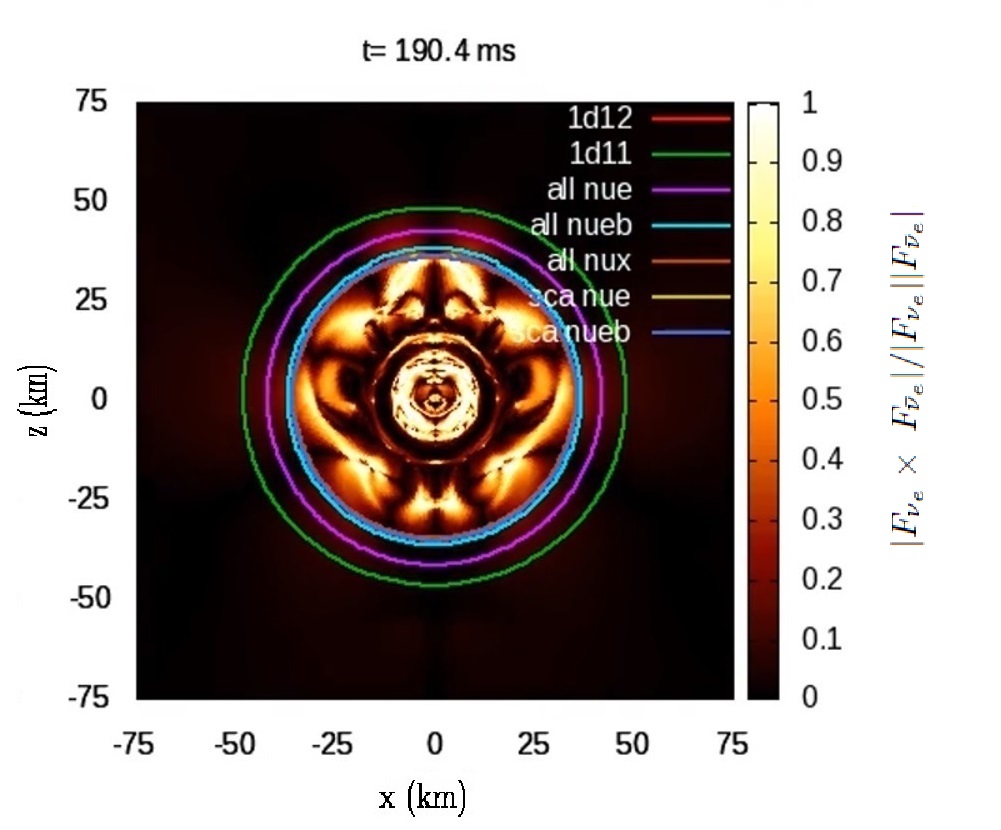} \\
\includegraphics[width=6.1cm]{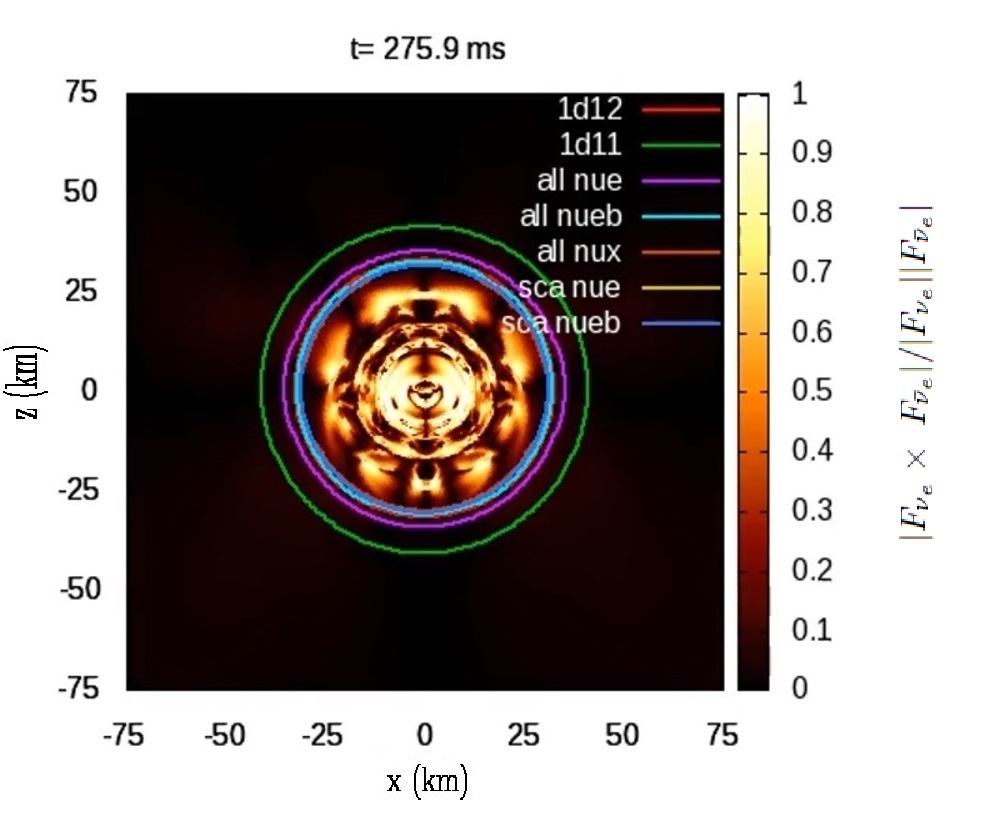} \\

\end{tabular} \caption{\label{275snap} The color contours for the sine of the angle that the flux vectors make: $|F_{\nu_{e}}\times~F_{\bar\nu_{e}}|/|F_{\nu_{e}}||F_{\bar\nu_{e}}|$.The red and green circles correspond to the densities of $10^{12}$ and $10^{11}$$\mathrm{g/cm^{3}}$, respectively. The violet, light-blue and brown circles indicate, respectively, the neutrino spheres of $\nu_e$, $\bar\nu_e$ and $\nu_x$, or the radii that their mean free paths calculated from both absorption and scattering rates are equal to the density scaleheight $\rho/(d\rho/dr)$. The yellow and dark-blue circles, on the other hand, show the neutrino spheres of $\nu_e$ and $\bar\nu_e$, respectively, calculated from the scattering rate alone.} 
\end{figure}

We shift our attention to the neutrino distributions at these three epochs now. Note first that they are the results obtained in the simulation that neglected possible neutrino oscillations entirely. Since the fast-pairwise conversion is supposed to feed on the difference in the angular distributions in momentum space between $\nu_{e}$ and $\bar\nu_{e}$, we show in Fig.~\ref{275snap} the sine of the angle between the flux vectors for $\nu_{e}$ and $\bar\nu_{e}$, which are meant to be a rough measure of misalignment in the angular distributions, as color contour plots at the same three post-bounce times. The energy-integrated flux vector of neutrino species $i$ is defined as
\begin{equation}
\textbf{\textit{F}}(\textbf{r}) = \int \frac{E^2~dE~d\mathbf {v}}{(2\pi)^3}~{\mathbf {v}} f_{i}({\mathbf{r}}, E, {\mathbf{v}}).  
\end{equation}
Note that brighter colors indicate that the two flux vectors are highly misaligned. In the same figure we also give the iso-density surfaces for $\rho=10^{11},10^{12}$$\mathrm{g/cm^{3}}$ as well as the neutrino spheres for all species.

 As mentioned earlier, the convective motion occurs and the matter distribution is non-spherical in the PNS ($\rho\gtrsim10^{12}$$\mathrm{g/cm^{3}}$) and, as a result, neutrinos are flowing non-radially in the laboratory frame. Note that it is important to treat relativistic aberration properly in the neutrino transfer calculation in order to get the correct angular distribution in the laboratory frame in the optically thick region \cite{2018ApJ...854..136N}. More importantly, the flux vectors for $\nu_{e}$ and $\bar\nu_{e}$ are highly misaligned there as should be clear from the figure. Since in the linear analysis formulated in the previous section we ignored all neutrino interactions other than the forward-scatterings that induce the refractive effect, it is not applicable in principle to the region deep inside the neutrino sphere, where such neglected interactions may not be ignored actually (but see also \cite{2018arXiv180806618C}). We hence pick up by inspecting the top panel of Fig.~\ref{275snap} a point ($r$ = 44.8 km,~$\theta$~=~2.36~radian) for linear analysis, which is close to the neutrino speheres and where we expect the highest misalignment of the flux vectors for $\nu_{e}$ and $\bar\nu_{e}$.

\begin{figure}[h]
\begin{tabular}{c}
\includegraphics[width=5.5cm]{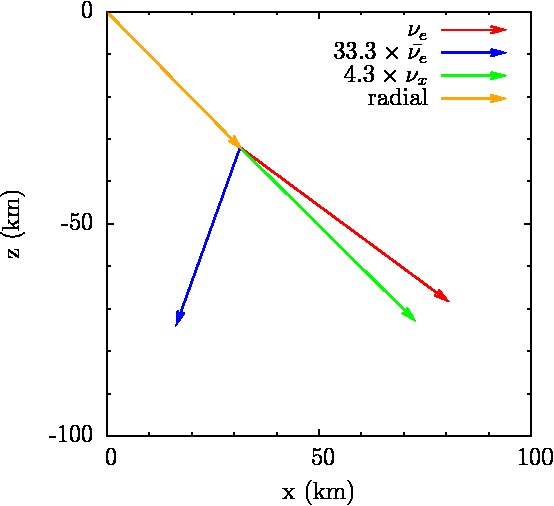}\\  
\includegraphics[width=5.5cm]{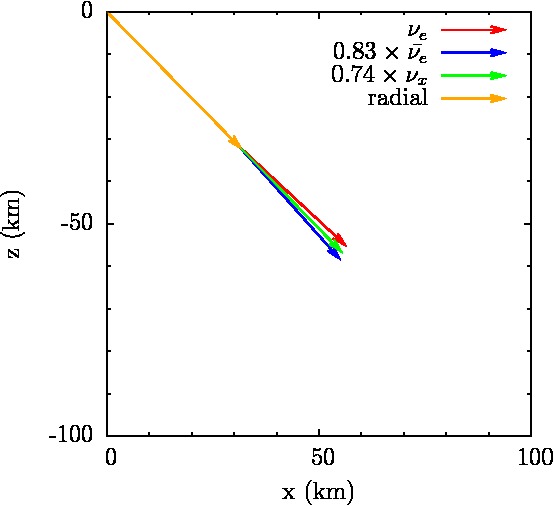}\\ 
\includegraphics[width=5.5cm]{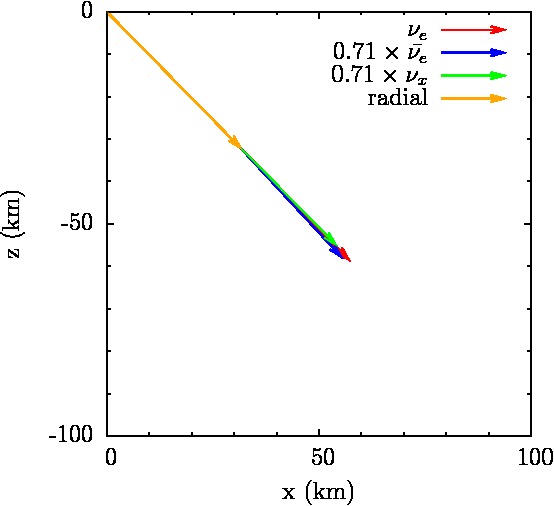} \\

\end{tabular} \caption{\label{flux2759} Energy-integrated fluxes of different neutrino species at the point indicated by the yellow radial lines ($r$~=~44.8~km,~$\theta$~=~2.36~radian). The top, middle and bottom  panels corresponds to $t_{pb}$~=~15.0~ms, $t_{pb}$ = 190.4 ms and \mbox{$t_{pb}$ = 275.9 ms, respectively.}}
\end{figure}

In order to see the typical time evolution, we employ the neutrino distributions at the same point for the later times, i.e., $t_{pb}$~=~190.4~ms and $t_{pb}$~=~275.9~ms although the misalignments are much reduced at this point for these times. This is confirmed indeed in Fig.~\ref{flux2759}, in which we exhibited the flux vectors for all three neutrinos.~As expected, the flux vectors for $\nu_{e}$ and $\bar\nu_{e}$ are almost perpendicular to each other at $t_{pb}$~=~15.0~ms.

In fact they are appreciably non-radial. This is because there is a convective motion of matter, to which they are coupled more strongly than~$\nu_x$. On the other hand, all three flux vectors are almost radial and hence aligned with one another at the later times as shown in the middle and bottom panels. It should be also noted that the absolute values of the fluxes are highly different among three species with the flux of $\bar\nu_{e}$ being the smallest in general as can be understood from the scaling factors employed in drawing the figure. This is simply because the Fermi degeneracy of electrons strongly supresses the population of $\bar\nu_{e}$. It is also evident that the asymmetry in abundance among neutrino species is somewhat relaxed as the time passes. As it will turn out later, these features have important implications for the fast-pairwise flavor conversion.

\begin{figure}[H]
\begin{tabular}{c}
\includegraphics[width=7cm]{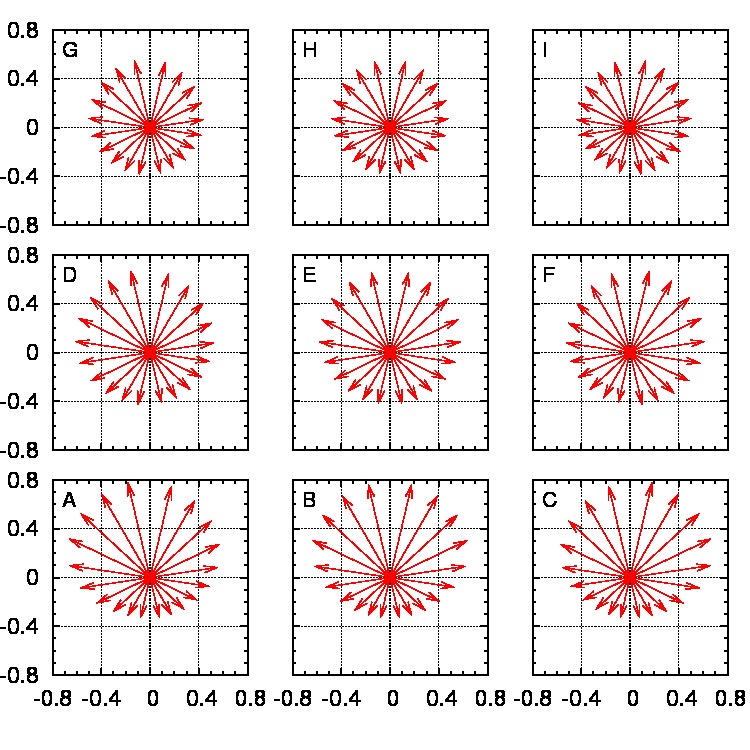}\\
\includegraphics[width=7cm]{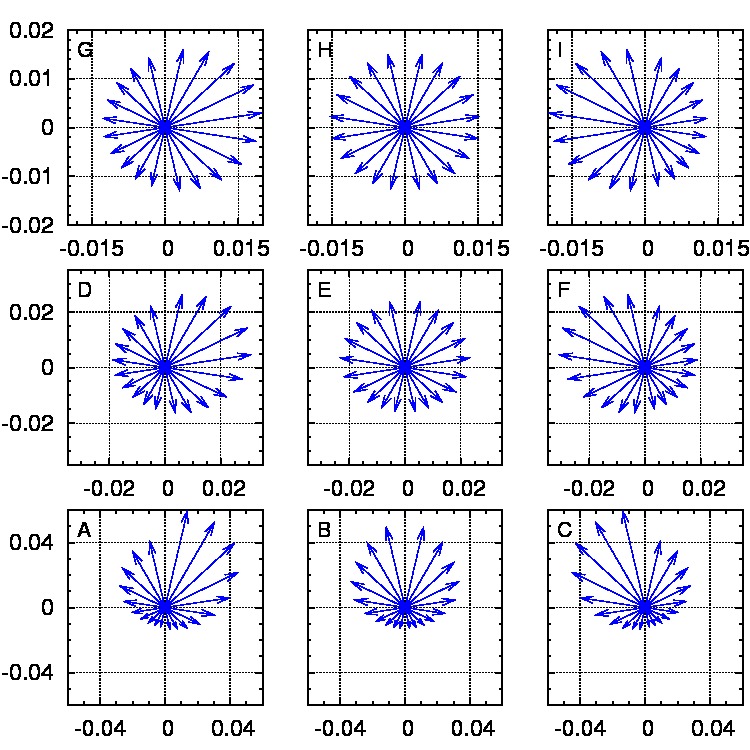}\\
\includegraphics[width=7cm]{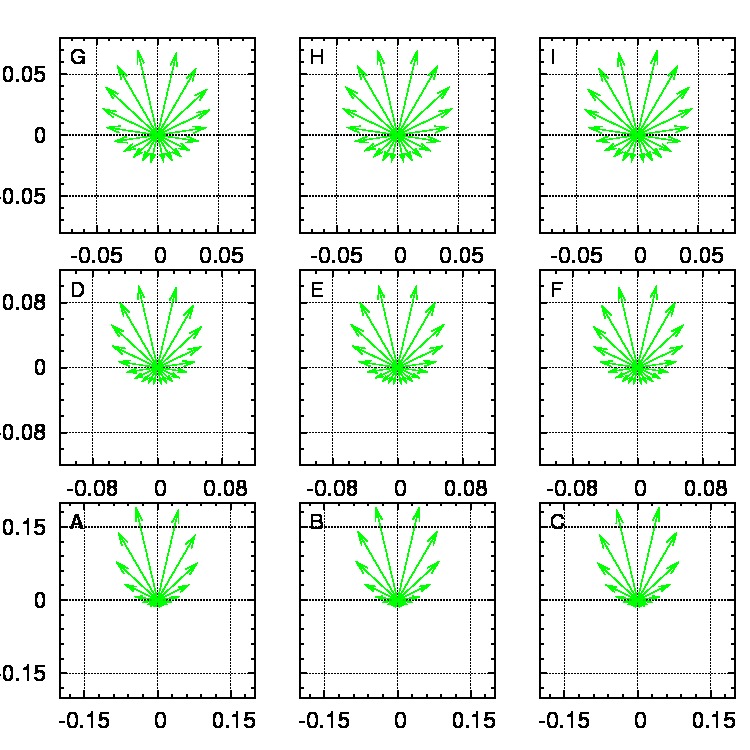}  

\end{tabular} \caption{\label{angnx15} Angular distributions in momentum space of $\nu_e$, $\bar\nu_e$ and $\nu_x$ at $t_{pb}$ = 15.0 ms. Panels A, B and C correspond to the neutrino energy of $E$~=~5~MeV whereas D, E and F have $E$~=~8.5 MeV and G, H and I share $E$ = 10 MeV. Each column shows the meridian sections, which correspond to a certain pair of $\phi_\nu$ values : $\phi_\nu$ [radian] = (0.35, 3.49), (1.57, 4.71) and (2.78, 5.92) for the left, middle and right columns, respectively.} 
\end{figure}

\begin{figure}[H]
\begin{tabular}{c}
\includegraphics[width=7cm]{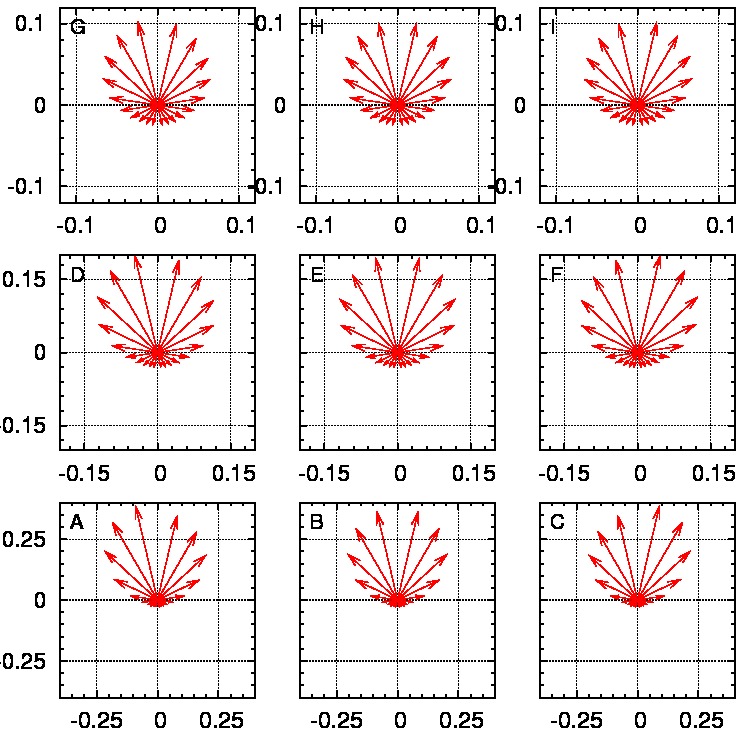}\\ 
\includegraphics[width=7cm]{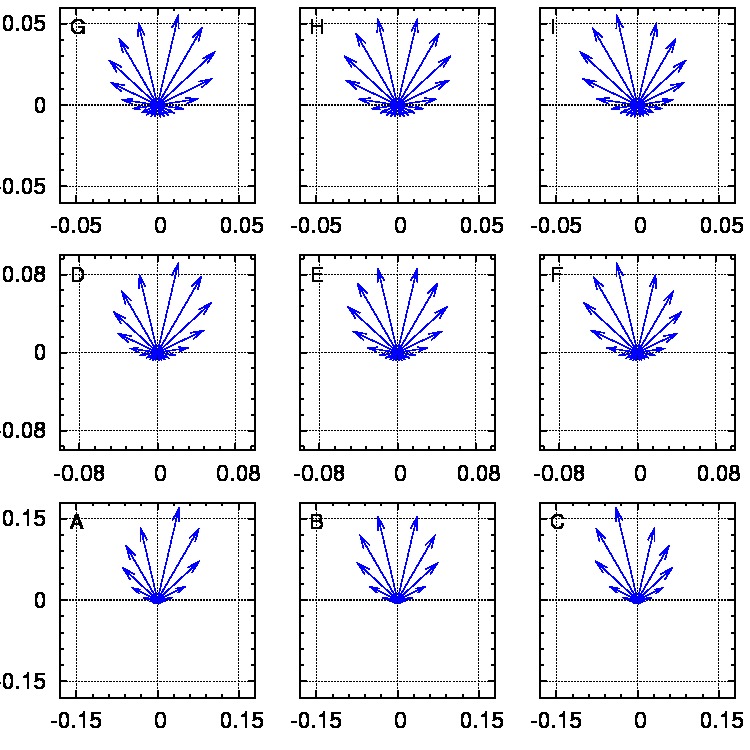}\\
\includegraphics[width=7cm]{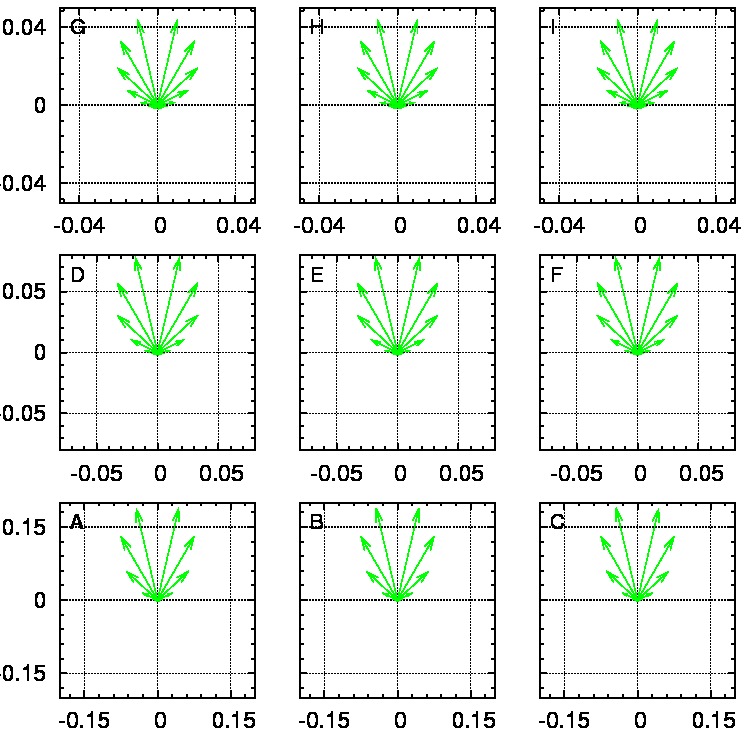}

\end{tabular} \caption{\label{angnx190} Same as Fig.~\ref{angnx15} but at $t_{pb}$ = 190.4 ms.} 
\end{figure}

\begin{figure}[H]
\begin{tabular}{c}
\includegraphics[width=7cm]{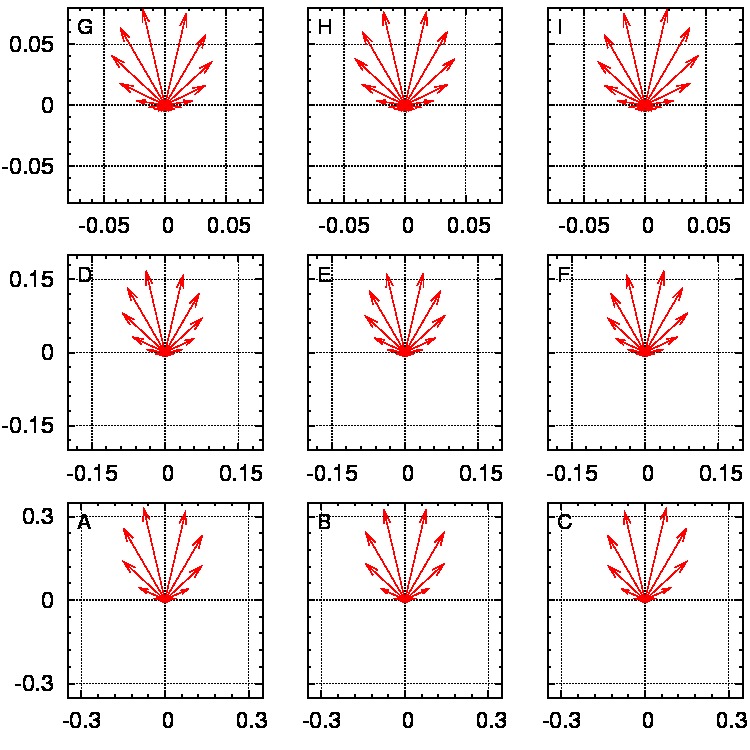}\\
\includegraphics[width=7cm]{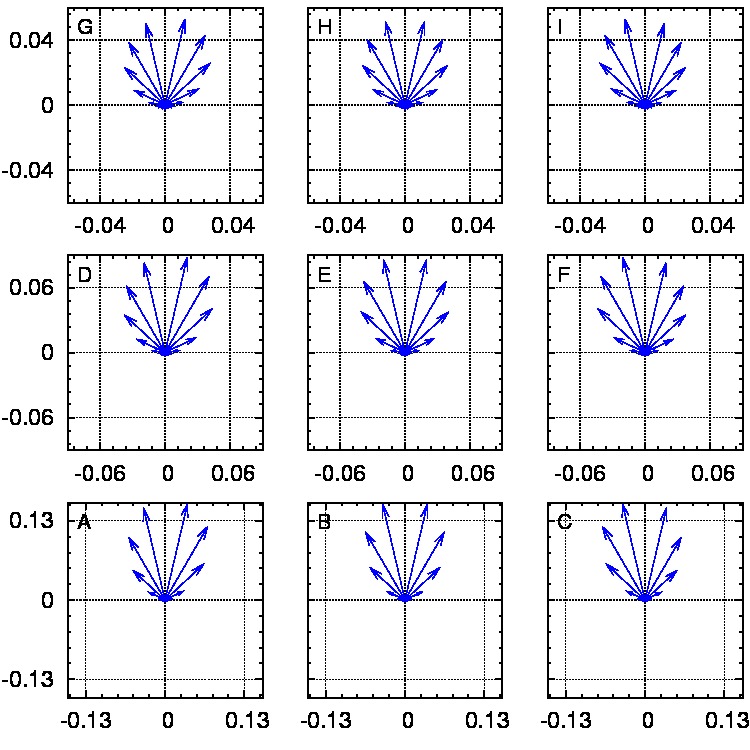}\\
\includegraphics[width=7cm]{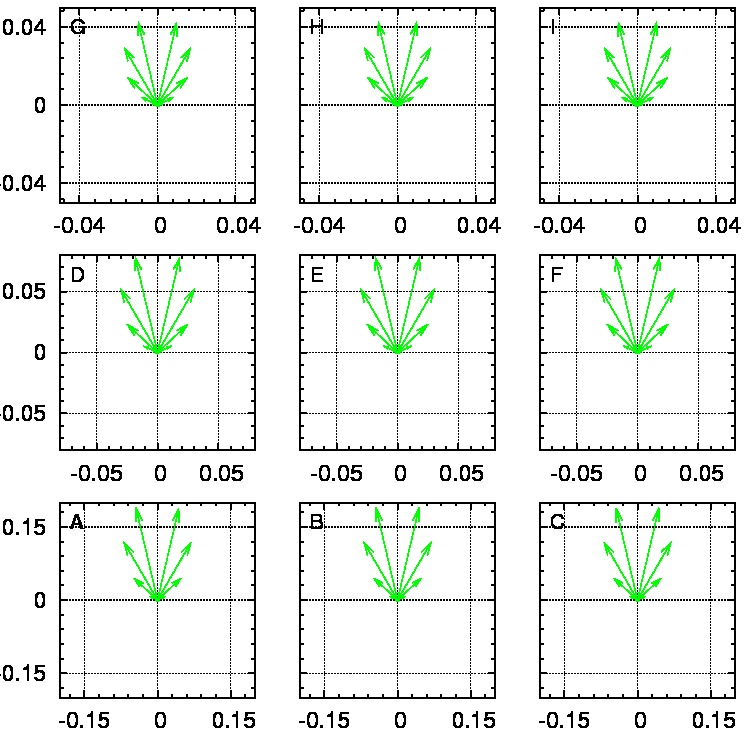}

\end{tabular} \caption{\label{angnx275} Same as Figs.~\ref{angnx15} and~\ref{angnx190} but at $t_{pb}$ = 275.9 ms.} 
\end{figure}

Figures~\ref{angnx15}, \ref{angnx190} and \ref{angnx275} show the angular distributions in momentum space of $\nu_{e}$, $\bar\nu_{e}$ and $\nu_{x}$ at the same spatial points for \mbox{$t_{pb}$ = 15.0}, 190.4 and 275.9 ms, respectively. They are actually the raw data extracted from the simulation and are the angular distributions of the neutrinos with specified energies in the meridian sections of momentum space at different values of $\phi_{\nu}$, which is the azimuthal angle in momentum space and should not be confused with the spatial one, $\phi$. Each arrow indicates the value of the neutrino distribution function in these figures in one of the $\theta_{\nu}$ bins adopted in the simulation. Again $\theta_{\nu}$ is the zenith angle in momentum space measured from the radial direction \cite{2018ApJ...854..136N} and should not be confused with the spatial counterpart $\theta$. The vertical axis corresponds to the radial direction at this point. The colors indicate neutrino species: the red, blue and green colors correspond to $\nu_e$, $\bar\nu_e$ and $\nu_x$, respectively. 

In each set of nine panels with one of the colors, three panels in a row have an identical neutrino energy while those in each column share the same value of $\phi_{\nu}$. Hence if one looks at these panels horizontally, one sees the azimuthal dependence whereas the energy dependence is found if they are viewed vertically. More specifically, the bottom row or panels A, B and C correspond to \mbox{$E$ = 5~MeV} and the middle row or panels D, E and F correspond to \mbox{$E$ = 8.5 MeV} and the top row or panels G, H and I share $E$ = 10 MeV. Note that the peak energy is $E$ = 8.5 MeV in the number spectrum of~$\nu_e$. These are actually the energies measured in the fluid rest frame in our simulation, which employs the two-energy grid \mbox{technique \cite{2014ApJS..214...16N}}, and, strictly speaking, we should \mbox{Lorentz-transform} them to the laboratory frame. Since the matter velocity is not very large, however, we ignore the slight differences that would make in the following analysis.
   
It is immediately apparent from the middle section with the blue color, i.e., for $\bar\nu_{e}$, that its angular distribution is not axisymmetric with respect to the local radial direction with the principal axis being inclined. This is also the case for other neutrino species although it is not so remarkable. Although $\nu_e$'s are most strongly coupled with matter, their flux is nearly radial accidentally. Anisotropy of the distribution function is more apparent for lower-energy neutrinos, since they are decoupled from matter deeper inside.

As the time passes, the proto neutron star contracts owing to neutrino emissions. As a result, the neutrino spheres also retreat to smaller radii and the neutrino angular distributions become more forward-peaked with inward-going neutrinos getting scarce.~This is clearly seen in Figs.~\ref{angnx190} and \ref{angnx275}.~As expected from the bottom panel of Fig.~\ref{angnx15}, the angular distributions are almost radially directed for all three neutrino species although some inclinations to the local radial direction are still visible for $\nu_e$ and $\bar\nu_{e}$ particularly at low energies.

It should be reminded that the scales are different from panel to panel in these figures. In fact, $\bar\nu_{e}$ has the smallest populations in general and this is particularly the case at the earliest time as should be evident from the comparison of $\bar\nu_e$ (blue sections) in Figs.~\ref{angnx15}, \ref{angnx190} and \ref{angnx275}. As mentioned earlier, this happens because the \mbox{Fermi degenracy} of electrons and, as a result, of $\nu_e$ as well is strong at the position of our current concern. It is found even at the latest time $t$ = 275.9 ms that the abundance of $\bar\nu_{e}$ is still less than half that of $\nu_e$ around their average energy. This disparity between $\nu_e$ and $\bar\nu_{e}$ is indeed unfavorable for the fast-pairwise flavor conversion as we will \mbox{demonstrate later.} 

\section{Results and Discussions}
\subsection{Investigations of the original data }

\begin{figure}[h]
\begin{tabular}{c}
\includegraphics[width=5.5cm]{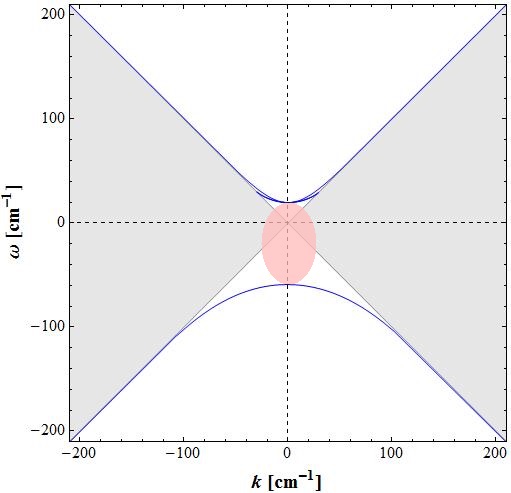}\\
\end{tabular}  \caption{\label{DRreal15} The dipersion relations between real $k$ and $\omega$ for the original data at $t_{pb}$ = 15.0 ms. It is assume that $\mathbf{k}$ is oriented in the crossing direction, which is different from case to case and is shown by an black arrow for each case in Figs. 10, 16 and 18. The red shaded area indicates the gap between branches, which is open in $\omega$ in this case.}

\end{figure}

We first look into the dispersion relations (DR's) between the wave number $k$ and the frequency $\omega$, which are obtained by applying the previous formulations to the realistic data for all three time steps. It should be emphasized here that the DR actually depends on the direction of $\mathbf k$. In fact our code can treat arbitrary directions \cite{2018PhRvD..97b3024M}. In the following, however, we mainly choose the direction, in which the \mbox{energy-integrated} \mbox{angular distributions} of $\nu_e$ and $\bar\nu_e$ are closest to each other and the crossing would occur if it could. We refer to it as the crossing direction hereafter. We will show later that as long as $\nu_e$ is dominant over $\bar\nu_e$, other directions of $\mathbf k$ normally give smaller growth rates if they are really unstable. In Fig.~\ref{DRreal15}, we show the DR's in blue for that direction of $\mathbf k$ and the gray region is the so-called zone of avoidance, i.e., an unphysical region. The red region indicates the gap between branches of the DR. There are multiple branches in general. They are stable modes, however, in which the amplitude does not grow in time. They are hence not interesting from a point of view of the flavor conversion. We then try to find complex $\omega$ solutions with positive imaginary parts, varying the value of real $k$, but in vain. It is hence highly likely that there is no such solution indeed for this case.

\begin{figure*}[t]
\begin{tabular}{ccc}
\includegraphics[width=6cm]{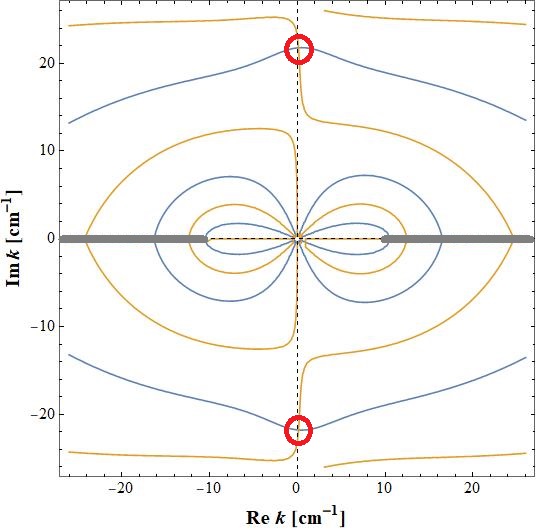} & \includegraphics[width=6cm]{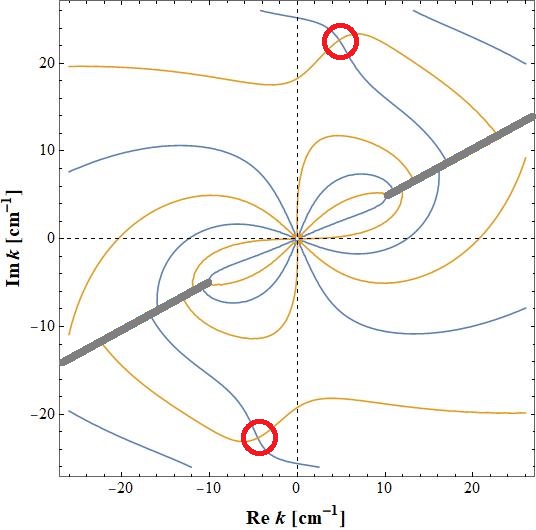} & \includegraphics[width=6cm]{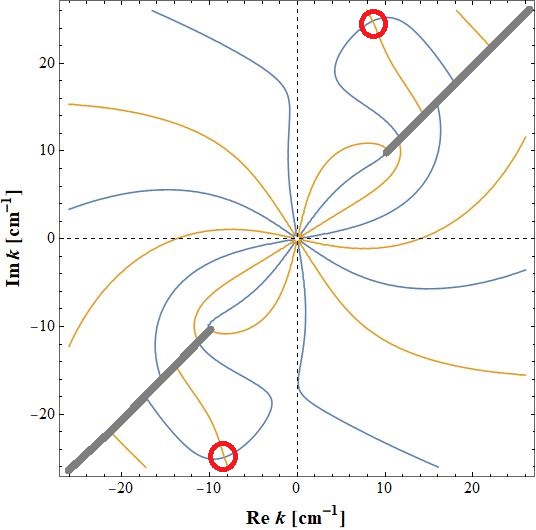}  \\
\includegraphics[width=6cm]{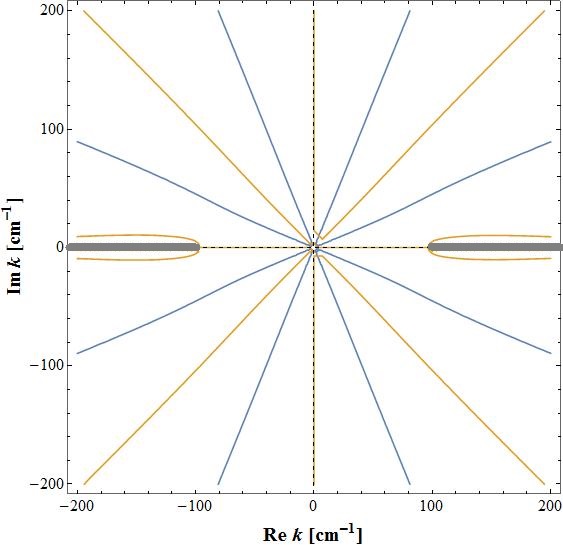} & \includegraphics[width=6cm]{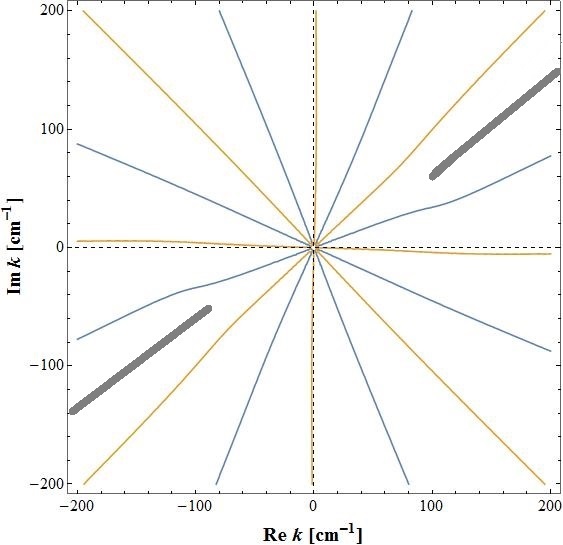} & \includegraphics[width=6cm]{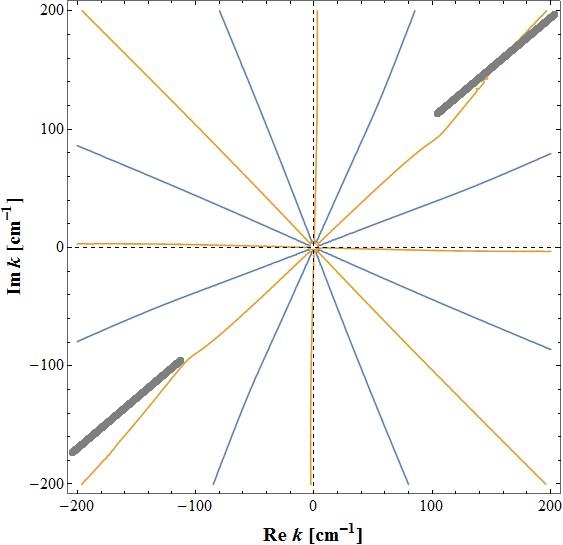}\\
\includegraphics[width=6cm]{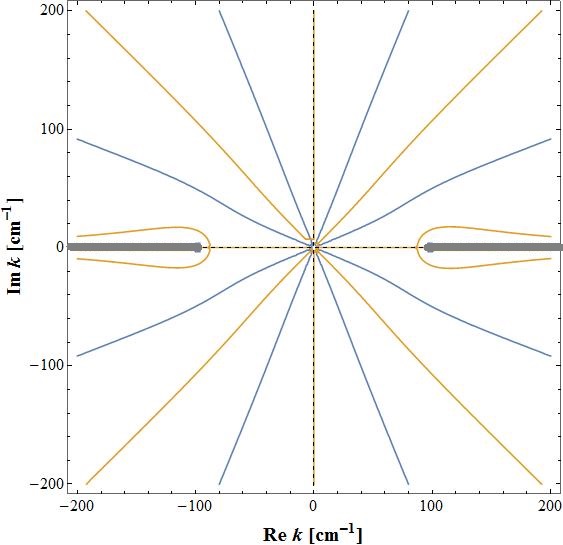} & \includegraphics[width=6cm]{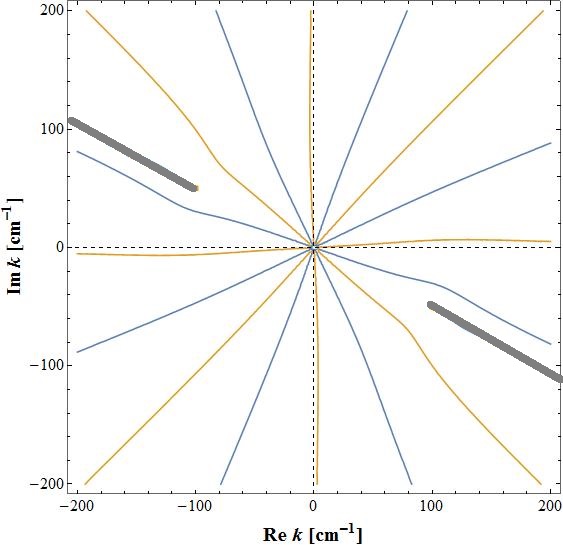} & \includegraphics[width=6cm]{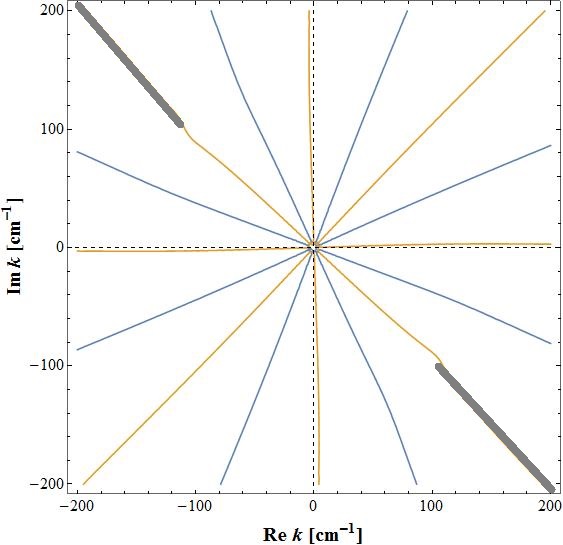}\\
\end{tabular}\caption{\label{DRroot15} Solutions of Eq.~(\ref{DR}) in $k$ for the realistic data at $t_{pb}$ = 15.0 ms. On the top row , the real part of $\omega$ is \mbox{Re $\omega=$ 10 $\mathrm{cm^{-1}}$} whereas its imaginary part is Im $\omega=$ 0, 5 and 10 $\mathrm{cm^{-1}}$ from left to right, respectively. The middle and bottom rows correspond to \mbox{Re $\omega=$ 100 and -100 $\mathrm{cm^{-1}}$}, respectively, and the imaginary part is Im $\omega=$ 0, 50 and 100 $\mathrm{cm^{-1}}$ for the left, middle and right columns, respectively. In each panel, blue and orange lines are the solutions of Re [det[$\Pi$]] = 0 and Im [det[$\Pi$]] = 0, respectively, and their intersections marked in red give actual solution of Eq.~(\ref{DR}). Note that we multiply $k$ with $\Pi$ in drawing these pictures and hence the origin is not a solution. The gray lines indicate the zone of avoidance.} 
\end{figure*}

In order to consolidate the claim, we look for solutions in another way. As mentioned above, looking at the DR's in Fig.~\ref{DRreal15}, one recognizes that there is a gap in $\omega$, i.e., the region that none of the branches traverses and is indicated in red. This implies that for any $\omega$ in this region there is no real $k$ that satisfies the Eq.~(\ref{DR}). We then expect that there are some complex $k$ solutions instead. This is exactly what we find in fact. Note again, however, that what we are seeking is not complex $k$ solutions for real values of $\omega$ but \mbox{complex $\omega$} solutions for real values of $k$. In order to see that these solutions do not exist, we take the following procedures.

We choose a couple of values of $\omega$ from the gap region and solve Eq.~(\ref{DR}) in the complex $k$-plane. As expected, we find some complex solutions in general as we will demonstrate shortly. We then gradually increase the imaginary part of $\omega$ from zero and solve Eq.~(\ref{DR}) repeatedly to see how the complex $k$ solutions move in the complex $k$-plane. If one of them crosses the real axis at some point, that is the solution we are seeking for. If instead none of the solution approaches the real axis, it indicates that there is no such solution with a real $k$ for a complex $\omega$ with a positive imaginary part. 

We first choose $\omega=$ 10 $\mathrm{cm^{-1}}$, which is inside the gap, and solve Eq.~(\ref{DR}) in $k$. We show in Fig.~\ref{DRroot15} the solutions of Re~[det[$\Pi$]] = 0 and Im [det[$\Pi$]] = 0 as blue and orange lines, respectively, in the complex $k$-plane. The intersections of these lines actually give the solutions of Eq.~(\ref{DR}). The top left panel of Fig.~\ref{DRroot15} shows the solutions for this particular real $\omega$. As expected, there are complex solutions on the imaginary axis as marked with red circles. Note that one of the orange lines coincides with the real axis, since we choose the real value for $\omega$. It is also mentioned that we have multiplied $k$ with det[$\Pi$] in drawing this figure just for our convenience. This extra multiplication is the reason why some of the lines are emanating almost radially from the origin, $k$ = 0, which is certainly not a solution. Also shown with thick gray lines are the zone of avoidance, in which no physical solution can exist, for the current choice of $\omega$.

Now we see how these complex solutions will move in the complex $k$-plane by changing the imaginary part of $\omega$. Note that we are interested only in $\omega$ with positive imaginary parts. We show in the middle and right panels on the top row of the same figure the results for Im~$\omega$~=~5~and~10~$\mathrm{cm^{-1}}$, respectively. The real part is unchanged. As we can see, there is no indication that they approach the real $k$-axis, which means there is no instability in this case. Although we present only two cases here, we have actually tried many other values of the imaginary part and confirmed that this statement is true. This is also the case for other values of the real part of~$\omega$ in the gap.

\begin{figure*}[t]
\begin{tabular}{ccc}
\includegraphics[width=5cm]{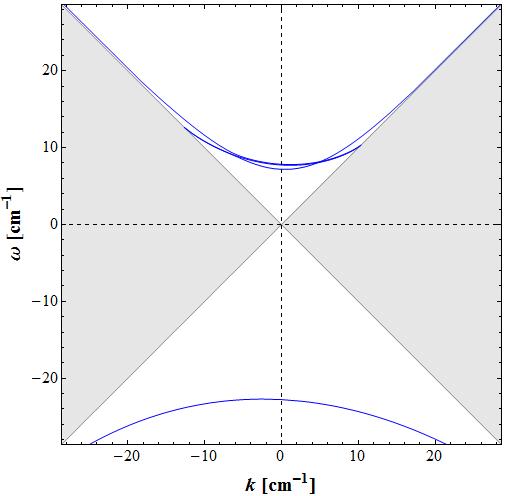} & \includegraphics[width=5cm]{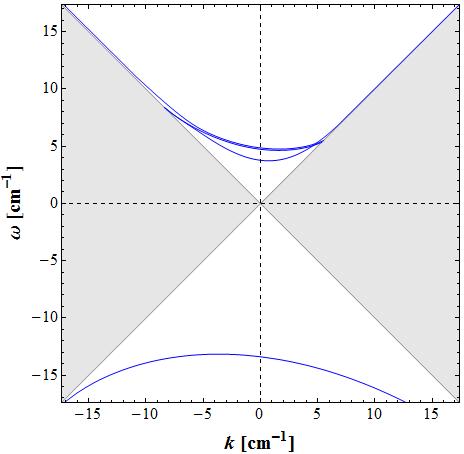} & \includegraphics[width=5cm]{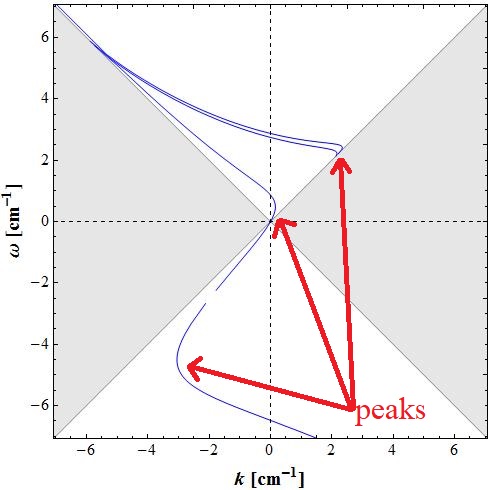} \\
\includegraphics[width=5cm]{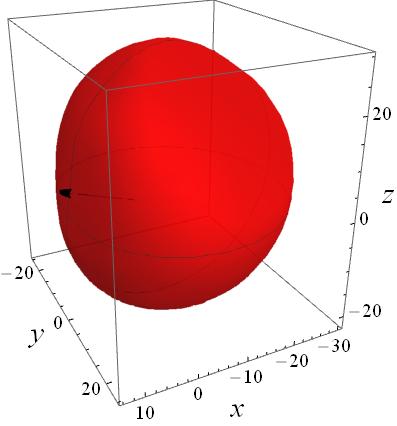} & \includegraphics[width=5cm]{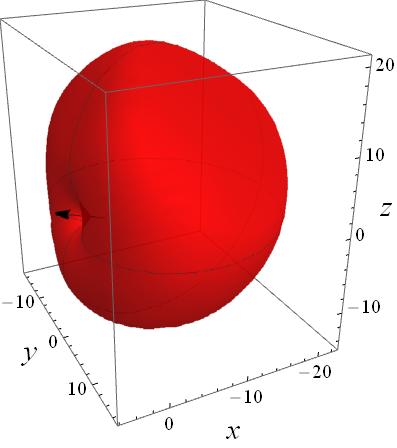} & \includegraphics[width=5cm]{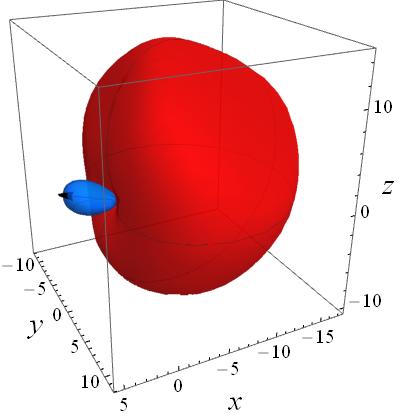}\\
 \\
\includegraphics[width=5cm]{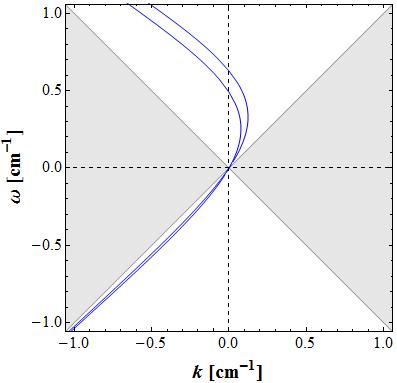} & \includegraphics[width=5cm]{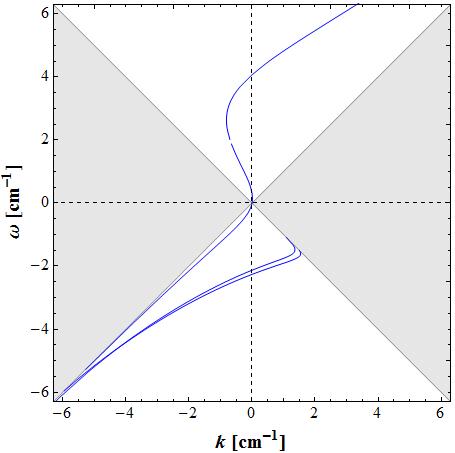} & \includegraphics[width=5cm]{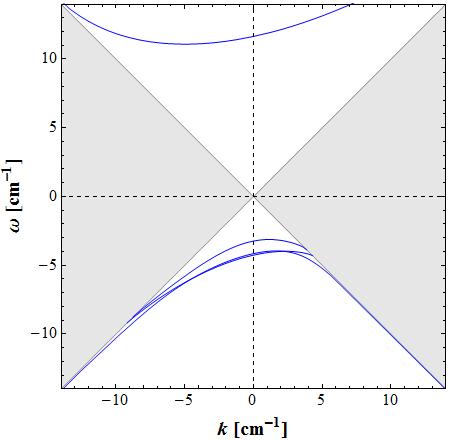}\\
\includegraphics[width=5cm]{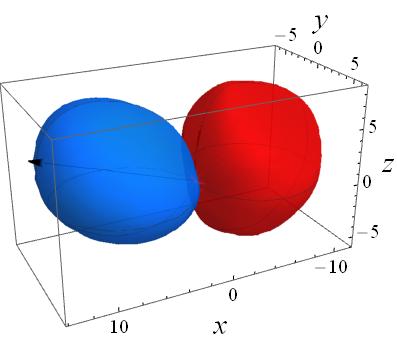} &\includegraphics[width=5cm]{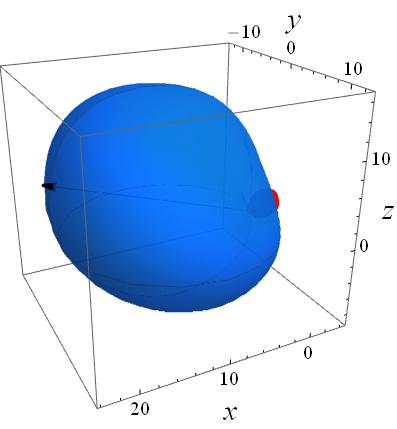} & \includegraphics[width=5cm]{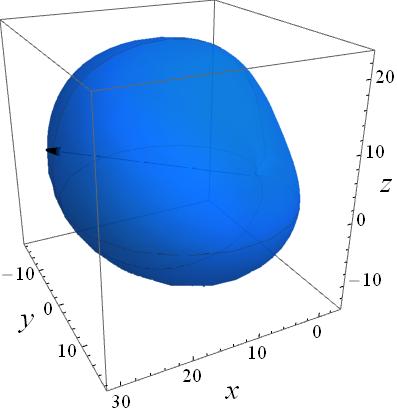}\\ 
 
\end{tabular}\caption{\label{DR-Rend15} The dispersion relations (first and third rows) and the angular distribution differences (second and fourth rows) between $\nu_e$ and $\bar\nu_e$ for modified data at $t_{pb}$ = 15.0 ms. We multiply the original distribution functions of $\bar\nu_e$ by a factor 25, 31 and 35, respectively, on the second row from left to right whereas the multiplication factor is 39.5, 45 and 49, respectively, on the fourth row from left to right. In these pictures, the angular distribution differences are expressed as colored surfaces. The red color implies that $\nu_e$ is dominant while the blue hue means otherwise. The distance from the origin to a point on the surface is equal to the absolute value of the difefrence for the direction to the point. The coordinates are the same as those deployed locally in the simulation with the z-axis being aligned with the local raidal direction; the x-axis corresponds to $\phi_\nu=0$. Black arrows indicate the crossing direction, which coincides with the direction of $\mathbf k$  in these models. Note that the breaks found in some branches are just artifacts in drawing these pictures.} 
\end{figure*}

\begin{figure*}[t]
\begin{tabular}{ccc}
\includegraphics[width=6cm]{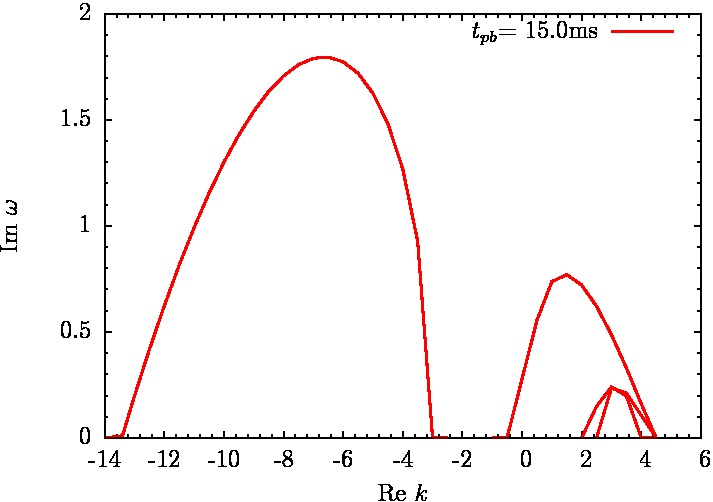} & \includegraphics[width=6cm]{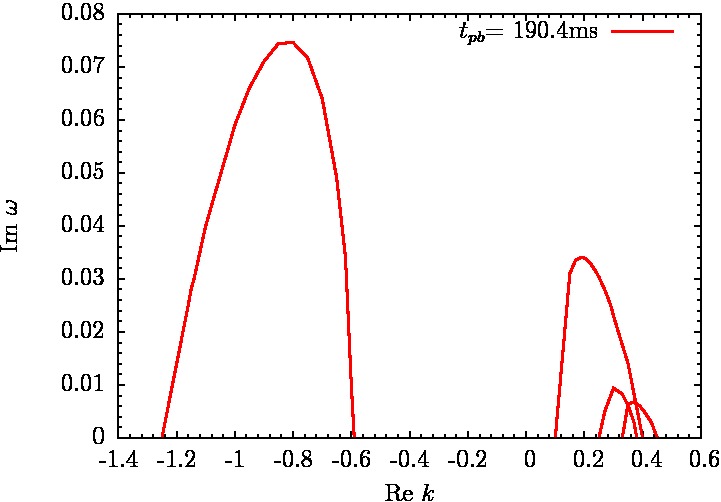} & \includegraphics[width=6cm]{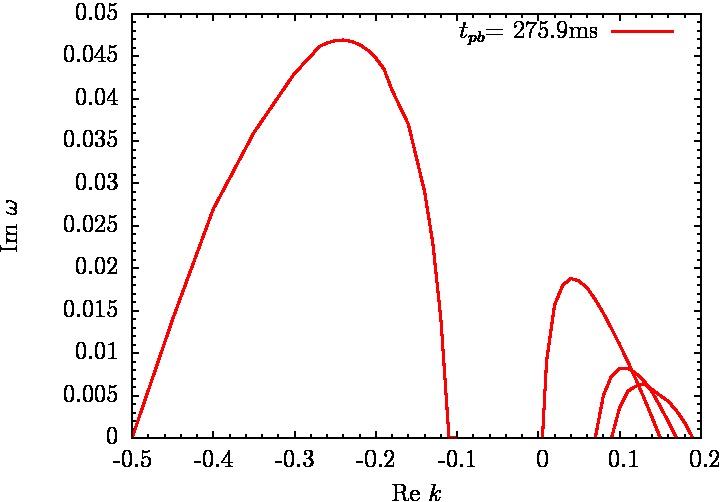}
\end{tabular}  \caption{\label{growth} The growth rate or Im $\omega$ as a function of $k$ for different time-steps when the crossing occurs for the first time. Note that $\mathbf k$ is oriented in crossing direction for all cases.}
\end{figure*}

\begin{figure*}[t]
\begin{tabular}{ccc}
\includegraphics[width=6cm]{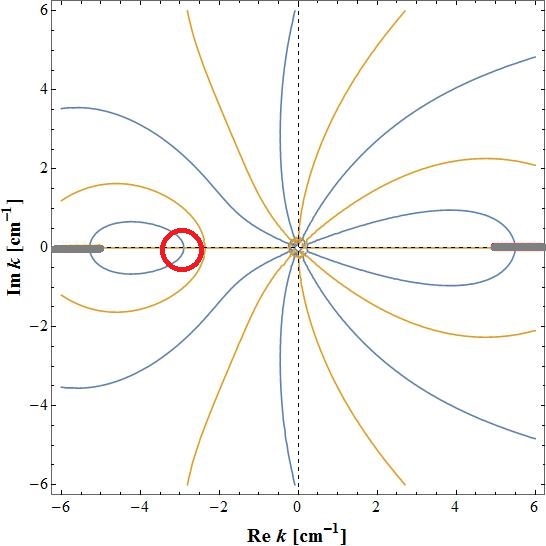} & \includegraphics[width=6cm]{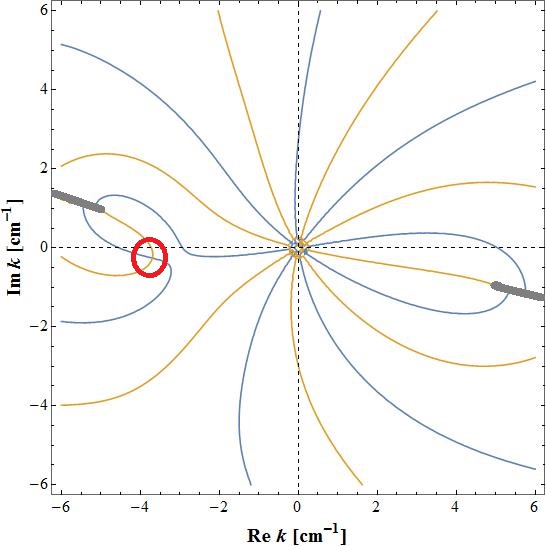} & \includegraphics[width=6cm]{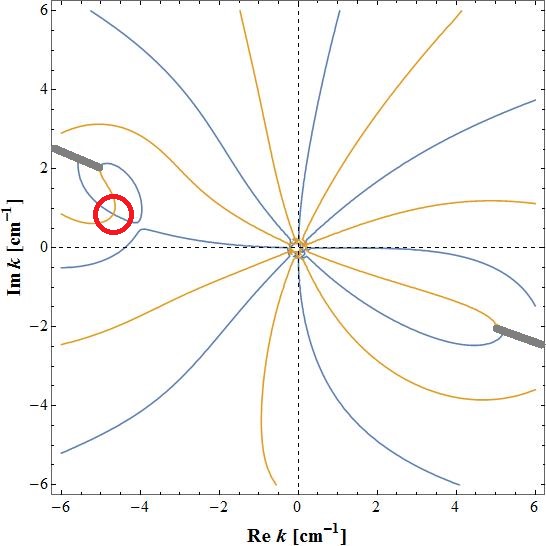}\\
 \end{tabular}\caption{\label{DRmod15} Solutions of Eq.~(\ref{DR}) in $k$ for the modified data at $t_{pb}$ = 15.0 ms. The distribution function of $\bar\nu_e$ is multiplied by a factor 35 (see the rightmost pictures on the first and second rows in Fig.~\ref{DR-Rend15}). We set Re $\omega=$ -5.0 $\mathrm{cm^{-1}}$ and Im $\omega=$ 0, 1.0 and 2.0 $\mathrm{cm^{-1}}$ from left to right. The red circles mark one of the solutions of det [$\Pi$] = 0 we focus here.}
\end{figure*}

\begin{figure*}[t]
\begin{tabular}{ccc}
\includegraphics[width=15cm]{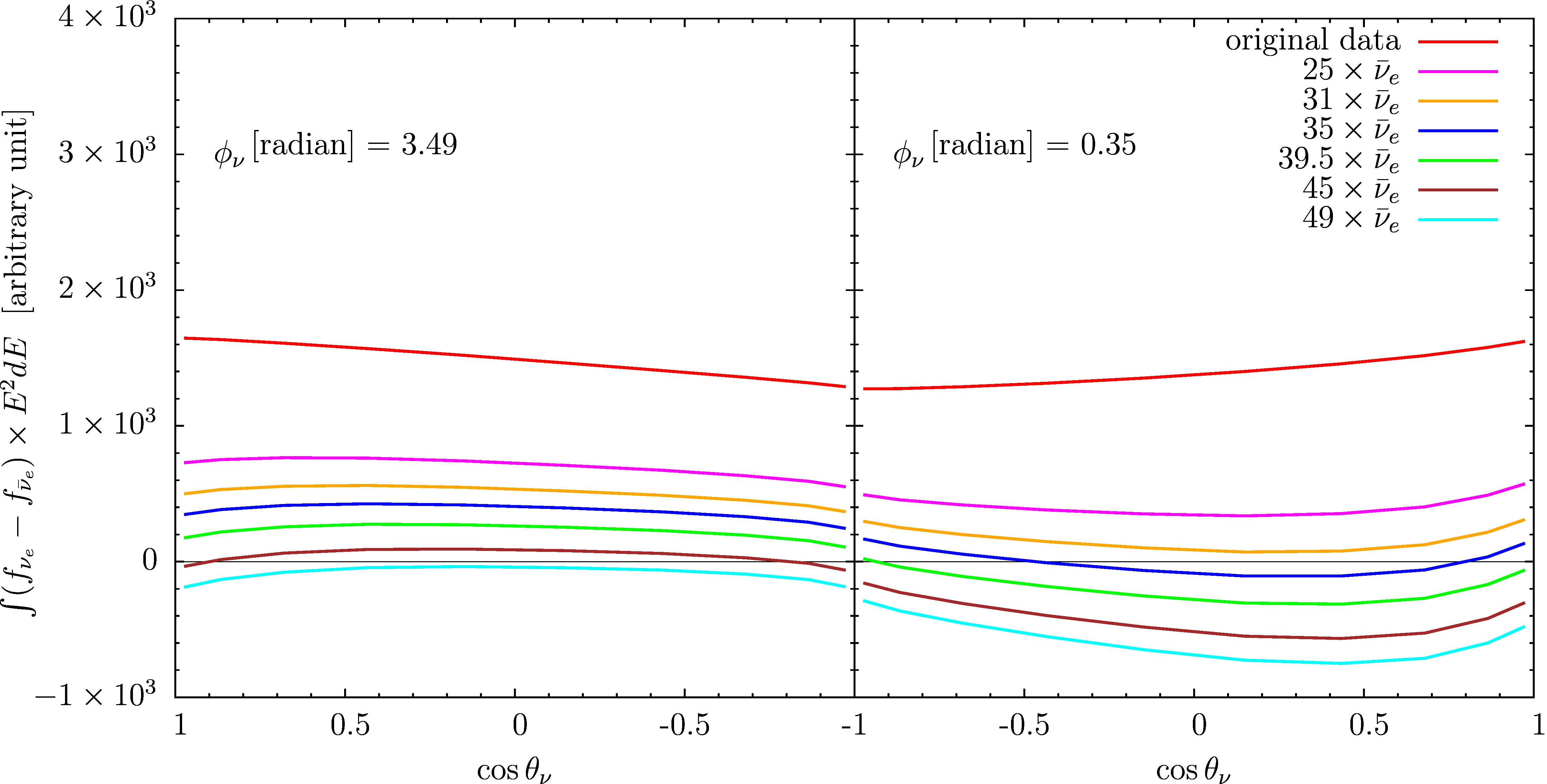}\\
 \end{tabular}\caption{\label{f_costheta} The angular-distribution difference between $\nu_e$ and $\bar{\nu}_e$ as a function of $\cos\theta_{\nu}$ for different multiplication factors at \mbox{$t_{pb}$ = 15.0 ms}. Right and left halves correspond to $\phi_{\nu}$ = 0.35 and 3.49 radian, or the first and fourth azimuthal-angle bins in the simulation, respectively. The crossing occurs for the first time between the multiplication factors 31 (orange line) and \mbox{35 (blue line).} }
\end{figure*}

\begin{figure*}[t]
\begin{tabular}{ccc}
\includegraphics[width=5cm]{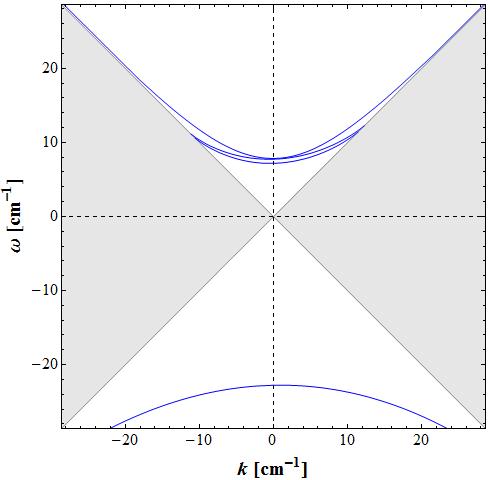} & \includegraphics[width=5cm]{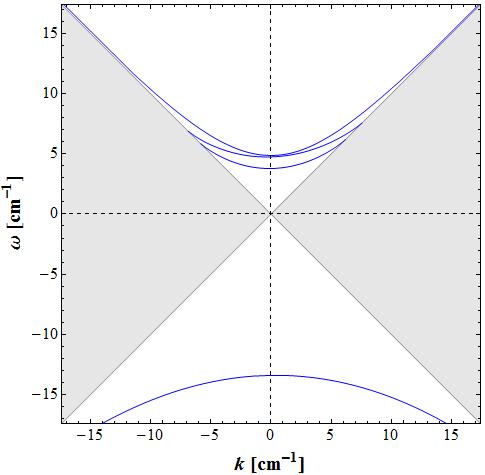} & \includegraphics[width=5cm]{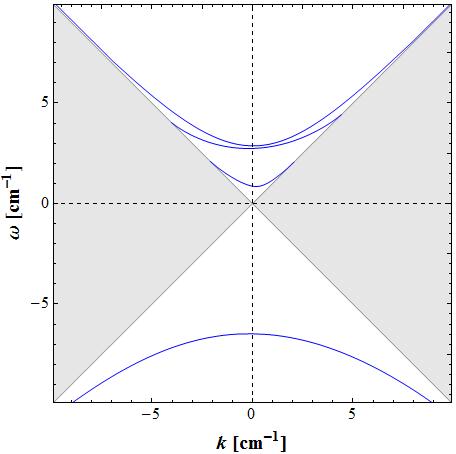}\\
\includegraphics[width=5cm]{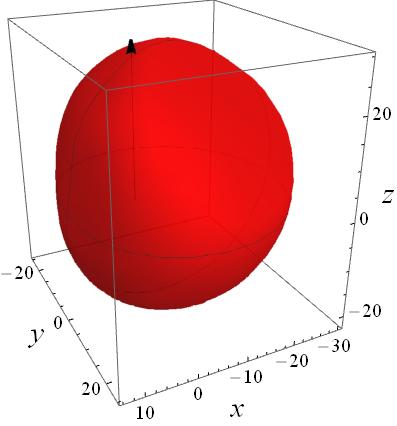} & \includegraphics[width=5cm]{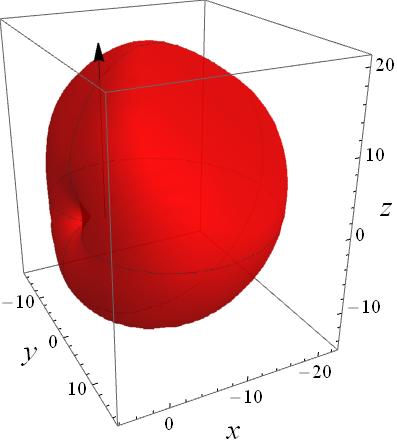} & \includegraphics[width=5cm]{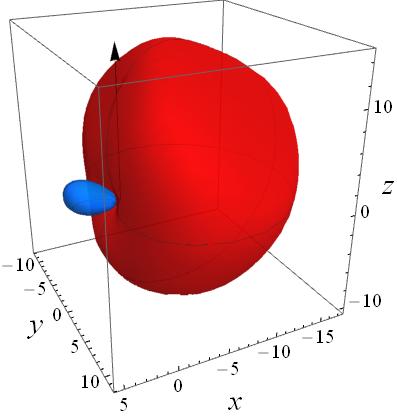} \\
\includegraphics[width=5cm]{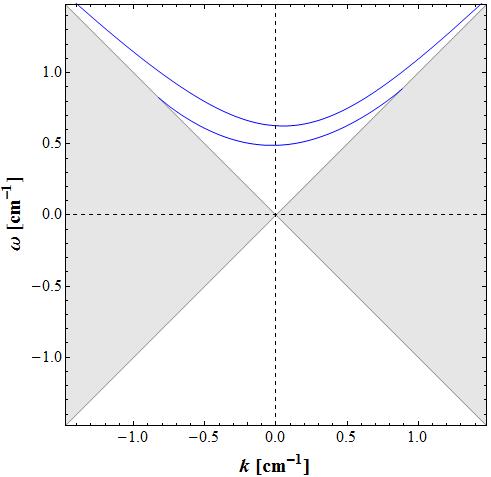} & \includegraphics[width=5cm]{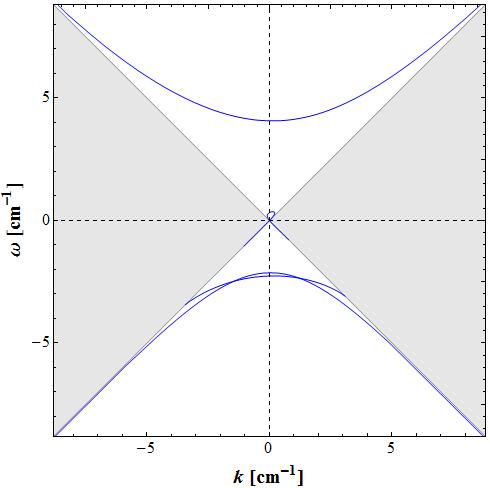} & \includegraphics[width=5cm]{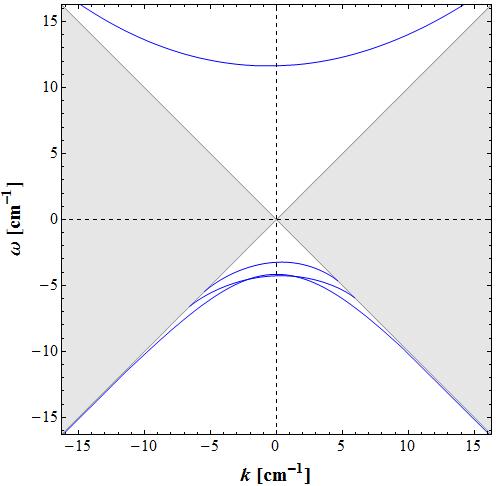}\\
\includegraphics[width=5cm]{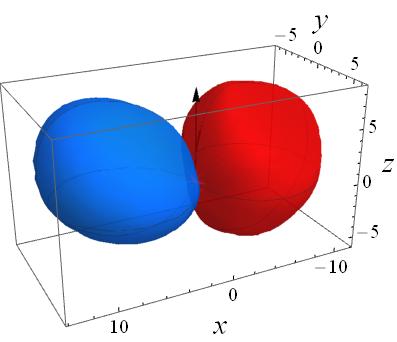} &\includegraphics[width=5cm]{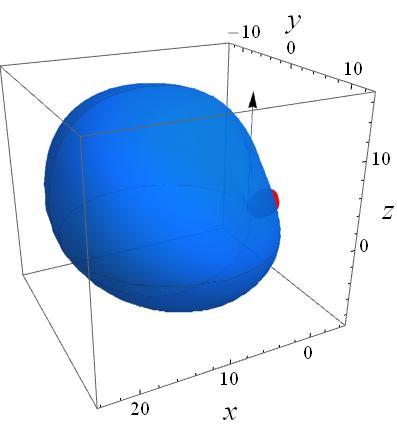} & \includegraphics[width=5cm]{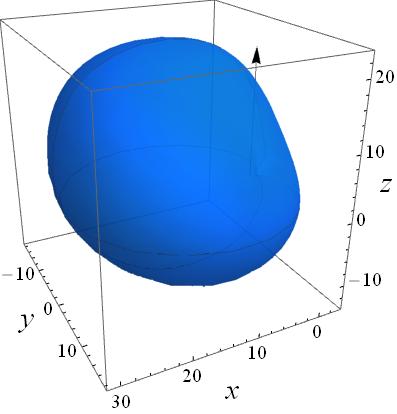}\\
 \end{tabular}\caption{\label{DR-Rend15-radial} Same as Fig.~\ref{DR-Rend15} but for $\mathbf k$ in the radial direction (or the positive z-direction).}
\end{figure*}

\begin{figure*}[t]
\begin{tabular}{cc}
\includegraphics[width=7cm]{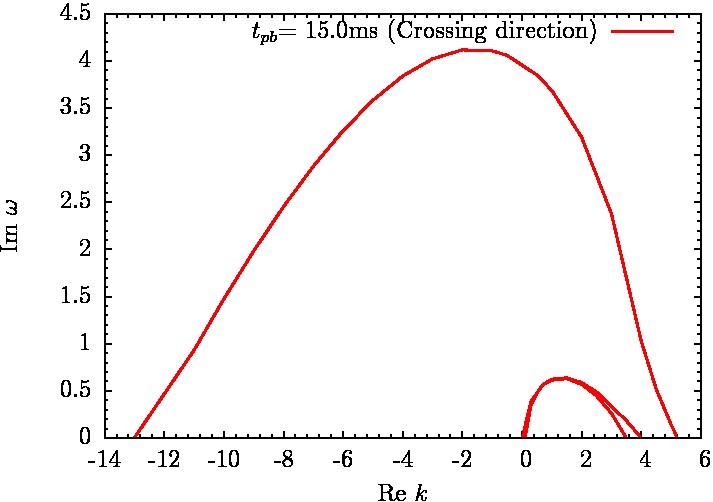} & \includegraphics[width=7cm]{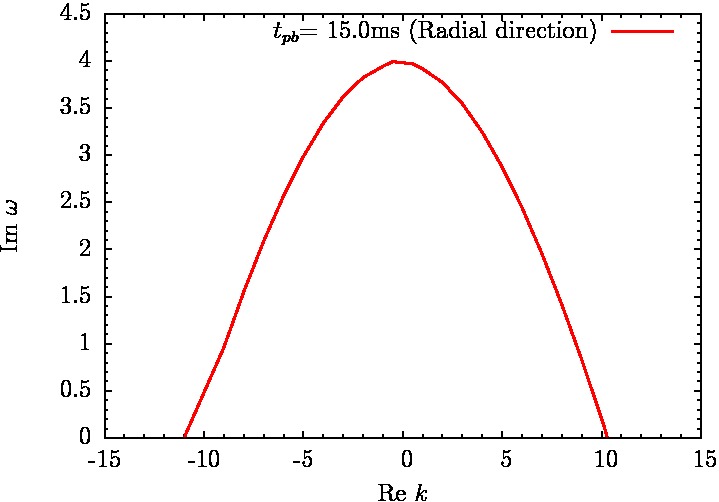}\\
 \end{tabular}\caption{\label{growth-new} A comparison of the growth rates as a function of $k$ between the crossing and radial directions at $t_{pb}$ = 15.0 ms. The multiplication factor is 39.5, which corresponds to the near equal populations of $\nu_e$ and $\bar{\nu}_e$.}
\end{figure*}

\begin{figure*}[t]
\begin{tabular}{ccc}
\includegraphics[width=6cm]{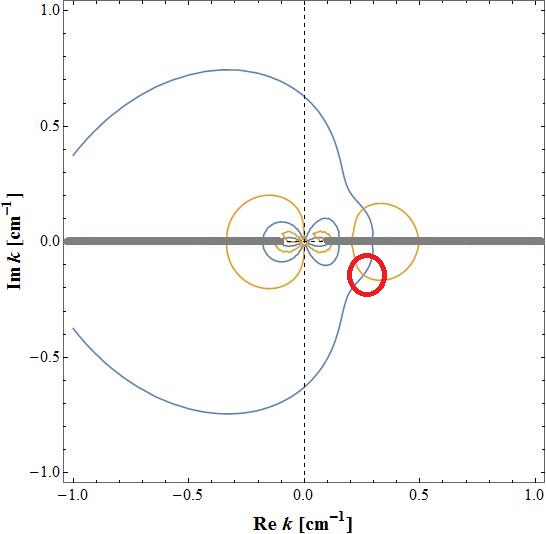} & \includegraphics[width=6cm]{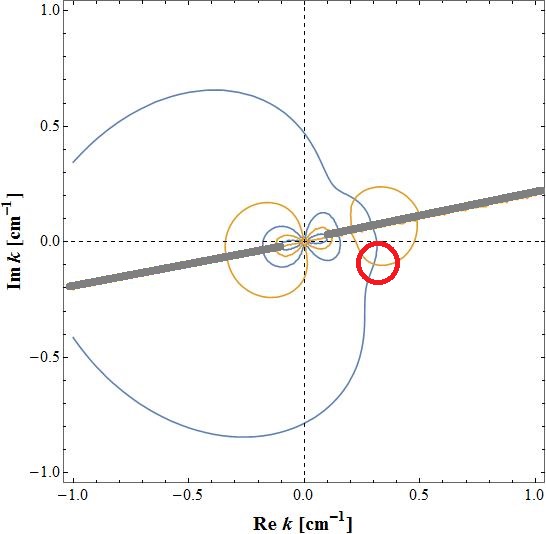} & \includegraphics[width=6cm]{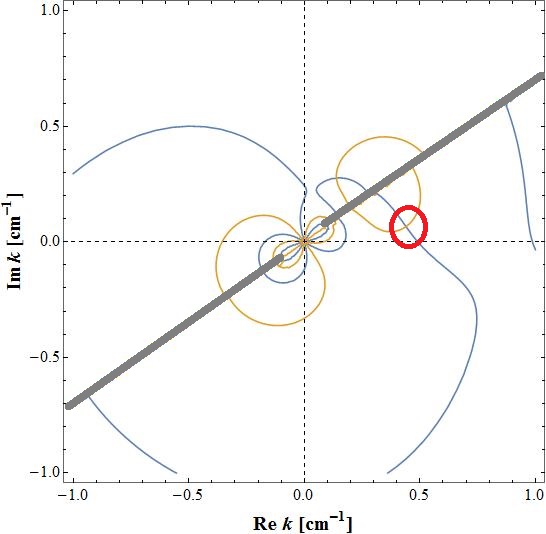}\\
 \end{tabular}\caption{\label{newkplane} Same as Fig.~\ref{DRmod15} but for $k$ in the radial direction. The distribution function of $\bar\nu_e$ is multiplied by a factor 35 (see the rightmost pictures on the first and second rows in Fig.~\ref{DR-Rend15-radial}). We set \mbox{Re $\omega=$~0.1 $\mathrm{cm^{-1}}$} and Im $\omega=$~0, 0.02 and 0.07 $\mathrm{cm^{-1}}$ from left to right. The red circle marks one of the solutions of det [$\Pi$] =~0 we focus here.}
\end{figure*}

We also investigate the regions outside the gap to check the possibility of the instability.~We first choose Re~$\omega$~=~100~$\mathrm{cm^{-1}}$ and set~Im~$\omega$~=~0, 50~and~100~$\mathrm{cm^{-1}}$. The second row of Fig. \ref{DRroot15} corresponds to these cases from left to right. As is clear and expected, we do not find any solutions in the complex $k$-plane for these values of $\omega$ except for those in the zone of avoidance on the real axis and hence there is no solution of Eq.~(\ref{DR}) that reaches the real $k$ axis, which would give rise to the instability. We then repeat the same analysis for another real value of $\omega$~=~-100~$\mathrm{cm^{-1}}$~outside~the~gap. We employ the same values of 0, 50 and 100 $\mathrm{cm^{-1}}$ for Im $\omega$. As shown in the bottom row of Fig.~\ref{DRroot15}, where again three panels correspond to these cases from left to right, respectively, there is no solution in the complex $k$-plane for these cases, either, and hence there is no chance of the instability.     

These analyses clearly endorse the claim that there is no unstable mode for this particular model. We have repeated the same analyses for the other two post-bounce times. We found no sign of instability either in these cases. Since the results are not much different from the one shown above for $t_{pb}$ = 15.0 ms, they are presented in Appendix. Having obtained these negative results, we change question instead of studying other places or times, which will be postponed to the forthcoming paper (Morinaga et al. in preparation): what would have been necessary then for the successful flavor conversion? We will address this issue in the following.

\subsection{Analysis of scaled data}

 The conventional idea is that the fast-pairwise flavor conversion needs a crossing in the angular distributions between $\nu_e$ and $\bar\nu_e$. As understood from Figs.~\ref{flux2759} and \ref{angnx15}, it was lacking in the original numerical data owing to rather large asymmetries in the populations. It should be stressed again, however, that it has yet to be demonstrated that the crossing in the angular distributions of $\nu_e$ and $\bar\nu_e$ is indeed the condition for the fast-pairwise conversion particularly for multi-dimensional settings as considered here. We will hence study it in the following, modifiying the original data by multiplying the distribution function of $\bar\nu_e$ with some factors so that they should have a similiar population to that of $\nu_e$ and repeat the same analysis to see if we could find the instability. 

\begin{figure*}[t]
\begin{tabular}{ccc}
\includegraphics[width=5cm]{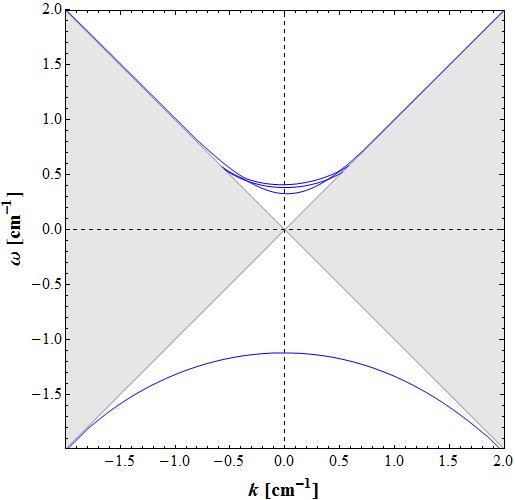} & \includegraphics[width=5cm]{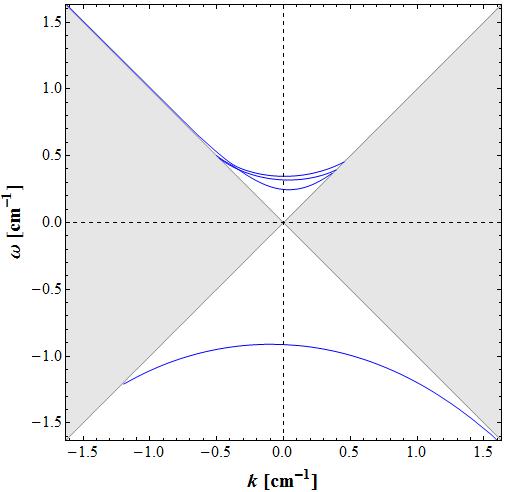} & \includegraphics[width=5cm]{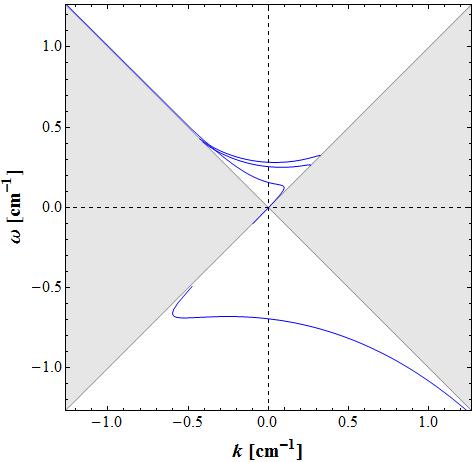}\\
\includegraphics[width=5cm]{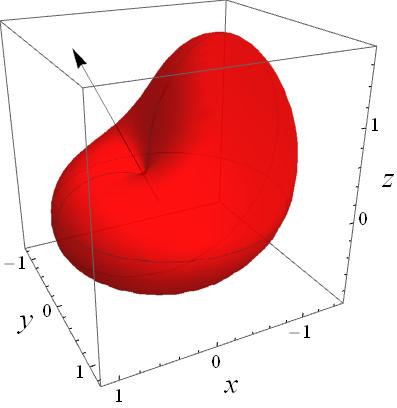} & \includegraphics[width=5cm]{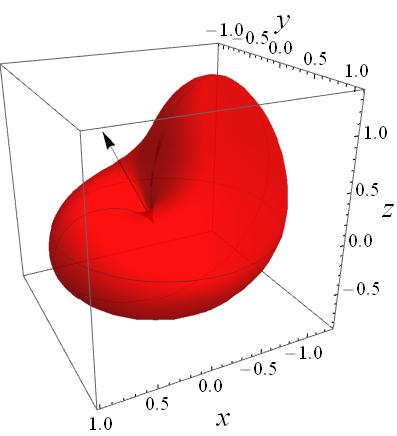}& \includegraphics[width=5cm]{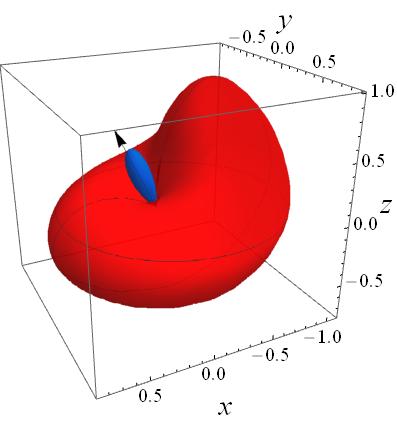}\\
\includegraphics[width=5cm]{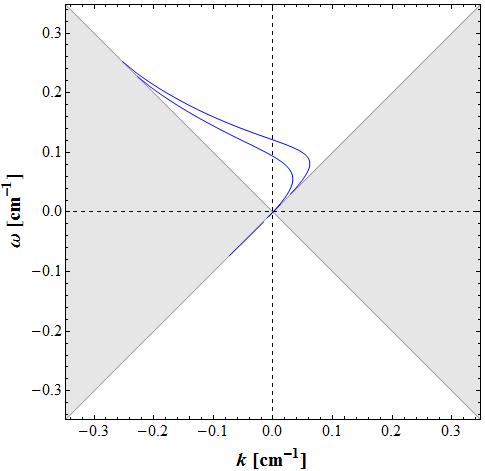} & \includegraphics[width=5cm]{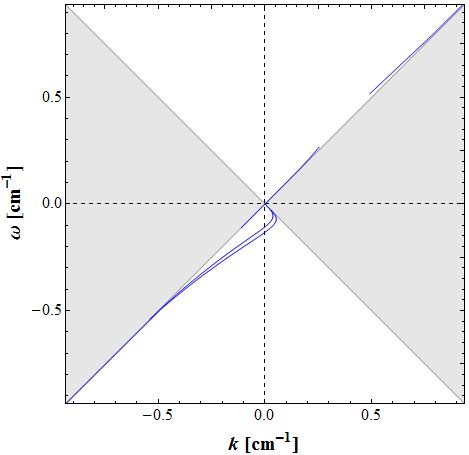} & \includegraphics[width=5cm]{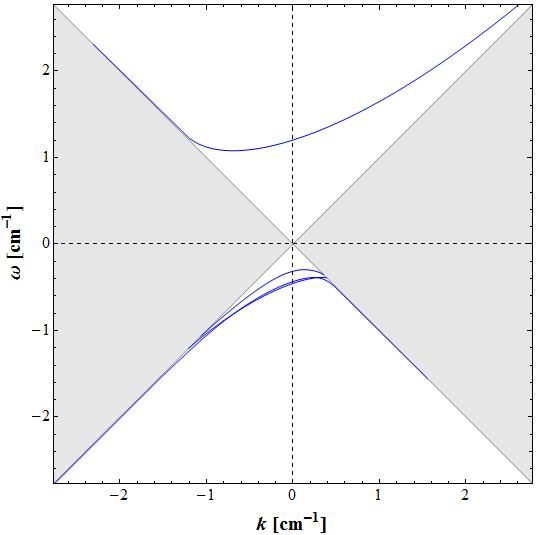}\\
\includegraphics[width=5cm]{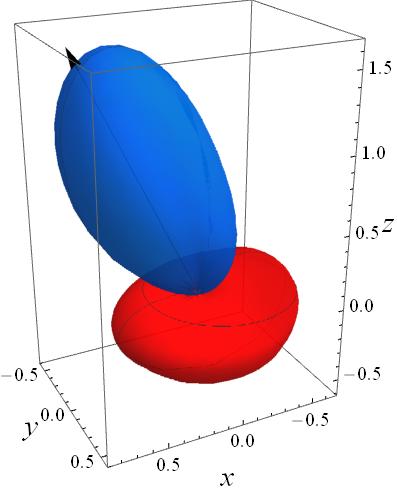} &\includegraphics[width=5cm]{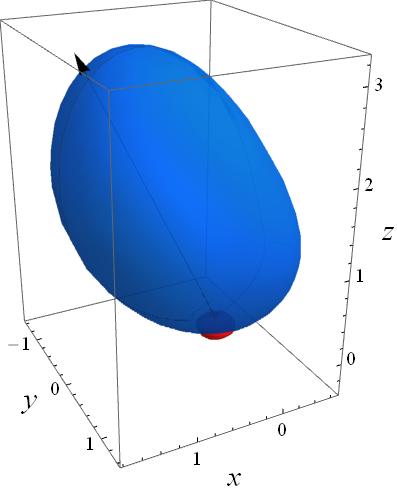} & \includegraphics[width=5cm]{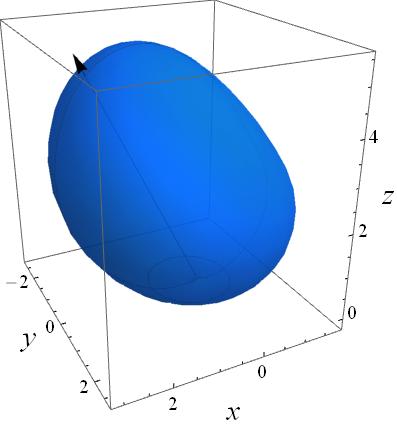}\\
 \end{tabular}\caption{\label{DR-Rend1904} Same as Fig.~\ref{DR-Rend15} but for modified data at $t_{pb}$ = 190.4 ms. The multiplication factor is 1.5, 1.6 and 1.7 for the top two rows while it is 1.95, 2.3 and 2.8 for the bottom two rows. Black arrows indicate the crossing direction in this case, to which we set the direction of $\mathbf{k}$.}
\end{figure*}

\begin{figure*}[t]
\begin{tabular}{ccc}
\includegraphics[width=6cm]{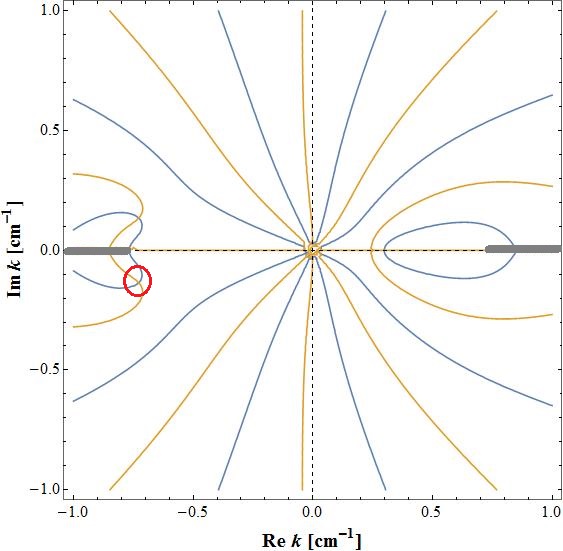} & \includegraphics[width=6cm]{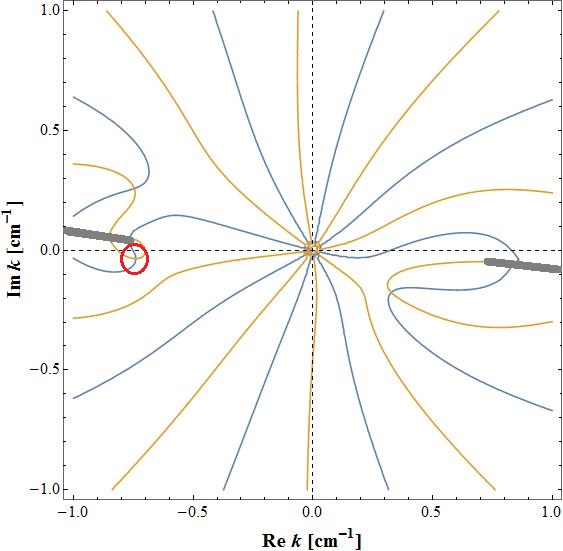} & \includegraphics[width=6cm]{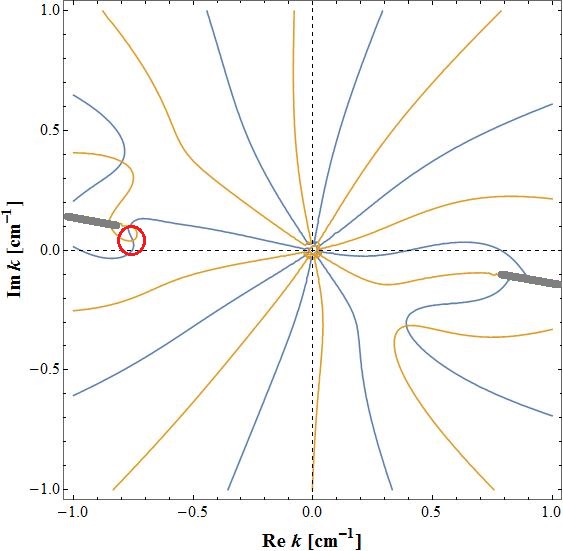}\\

\end{tabular}\caption{\label{DRmod1904} Same as Fig.~\ref{DRmod15} but at $t_{pb}$ = 190.4 ms. The direction of $\mathbf{k}$ is set to the crossing direction. The multiplication factor is 1.7 and \mbox{Re $\omega=$ -0.75 $\mathrm{cm^{-1}}$} and Im $\omega=$ 0.0, 0.05 and 0.1 $\mathrm{cm^{-1}}$ from left to right, respectively. The red circles mark one of the solutions of det [$\Pi$] = 0 we pay attention here.}
\end{figure*}

We first consider the case for $t_{pb}$ = 15.0 ms. As is clear from the top panel of Fig.~\ref{flux2759}, the fluxes of $\nu_e$ and $\bar\nu_e$ are almost perpendicular to each other but their magnitudes are also highly asymmetric in this case. We hence need to adopt a rather large factor $\sim$ 30 to obtain the crossing in the angular distributions.~Fig.~\ref{DR-Rend15} presents the the angular distribution differences between $\nu_e$ and $\bar\nu_e$ (second and fourth rows) as well as the corresponding DR's (first and third rows) for different multiplication factors. As we can see in the third panel from the left on the second row, the crossing occurs at about the multiplication factor of 35. Note that the crossing direction, which we have defined to be the direction, in which the crossing occurs for the first time, is almost aligned with the x-axis but is slightly inclined. In fact it makes 72 degree with the z-axis \mbox{($\theta_{\mathbf k}$ = 72 degree)}, which is actually the local radial direction. As mentioned earlier, this crossing direction is the direction of $\mathbf k$ we have chosen so far for the analysis of the DR at this time (see Fig.~\ref{DRreal15}). Note also that we choose different angles for other times as we will mention later. The third panel from the left on the first row indicates that the DR is qualitatively changed (c.f. Fig.~\ref{DRreal15}). There disappears the gap in $\omega$ and instead appear some peaks in $k$ as a function of $\omega$ near the borders with the zone of avoidance. This is an indication of the appearance of complex $k$ solutions in this case. 

This time we indeed find complex $\omega$ solutions for some ranges of real $k$. In Fig.~\ref{growth}, we show the growth rates, or the imaginary part of $\omega$, of these modes as a function of $k$. One can see that there are four unstable modes actually, corresponding to the peaks in $k$ in the DR. In order to see how the behavior of complex solutions changes from that for the original data, we investigate their movements in the complex $k$-plane by taking \mbox{Re~$\omega$=~-5.0~$\mathrm{cm^{-1}}$}, which corresponds to the lower left peak in the DR, and varying the imaginary part. This time there is a solution on the real axis, which is physical and located outside the zone of avoidance, as shown in the left panel of Fig.~\ref{DRmod15}. This is a stable mode we have found in the corresponding DR. This time this is the mode, whose movement we are interested in as we change Im $\omega$. The three panels of Fig.~\ref{DRmod15} correspond to the results for Im $\omega$ = 0, 1.0 and 2.0 $\mathrm{cm^{-1}}$, respectively. As one can clearly see from these plots, all the lines rotate clockwise but the solution of our concern marked with a red circle moves in the opposite direction initially and then goes upward and crosses the real $k$ axis somewhere between the last two panels. This corresponds to the unstable mode that gives the leftmost branch in the left panel of Fig.~\ref{growth}. One also finds that the maximum growth rate is attained near this point close to the lower left peak in the DR.

Now we look into the change in the behaviour of the DR more in detail. In order to demonstrate that this happens when the angular distributions start to have a crossing, we show in Fig.~\ref{DR-Rend15} the DR's and the angular distribution differences for other multiplication factors. The top two rows correspond to the multiplication factors of 25 and 31 and 35, respectively, whereas the bottom two rows show the results for the multiplication factors of 39.5, 45 and 49, respectively. As can be understood from the first two pictures, these factors are not large enough to produce the crossing. This is confirmed more clearly in Fig.~\ref{f_costheta}, in which we show the angular distribution difference between $\nu_e$ and $\bar{\nu}_e$ as a function of $\cos\theta_{\nu}$ for a pair of the values of $\phi_{\nu}$. The DR's are not different qualitatively from the original one. The crossing is indicated by a blue small side-lobe in rightmost picture on the second row.~The DR changes rather abruptly near that point. When the populations of $\nu_e$ and $\bar\nu_e$ become nearly the same (see the leftmost picture on the bottom row), the DR is changed qualititavely again. Finally in the bottom right panel, where $\bar\nu_e$ is much more abundant than $\nu_e$, which will be never realized in reality, the DR returns to something close to the one in the first two panels on the second row but reflected with respect to $k$ =~0, which should be as expected.  We confirm that there are unstable modes in all the cases with crossing and vice versa.

\begin{figure*}[t]
\begin{tabular}{ccc}
\includegraphics[width=5cm]{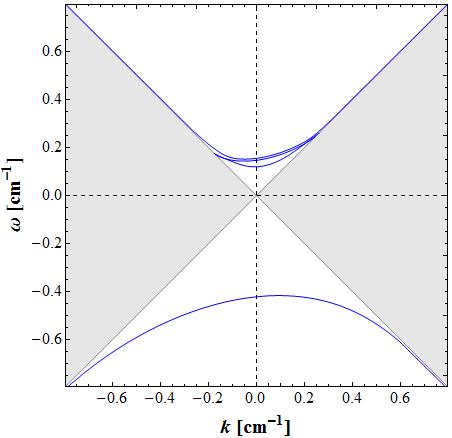} & \includegraphics[width=5cm]{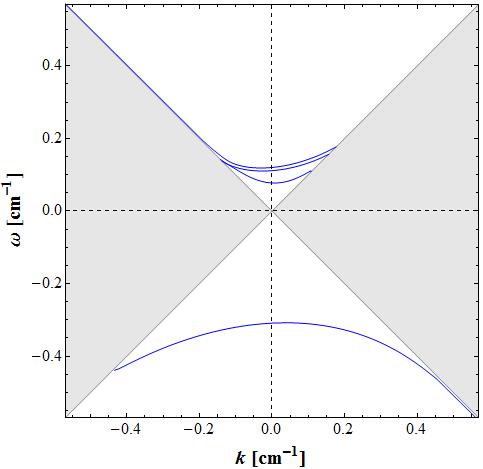} & \includegraphics[width=5cm]{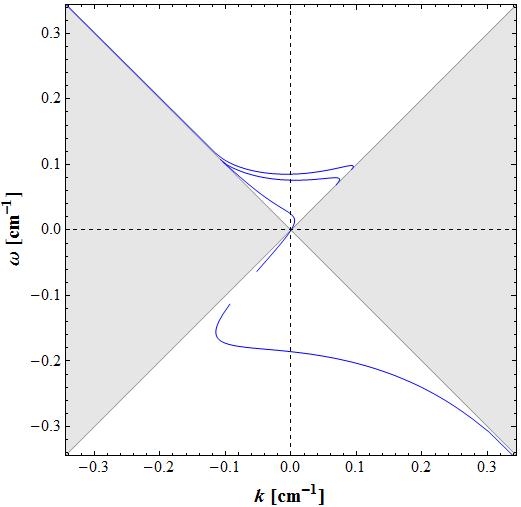}\\
\includegraphics[width=5cm]{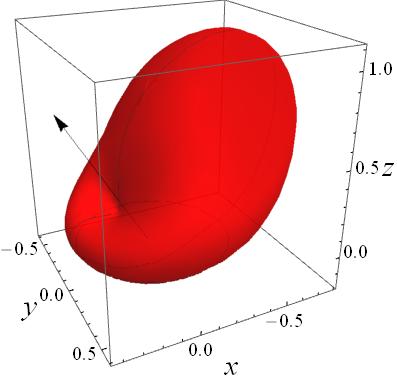} & \includegraphics[width=5cm]{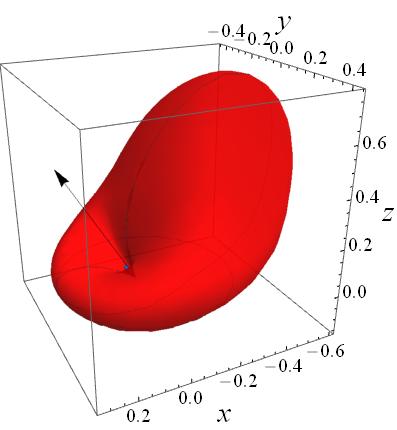} & \includegraphics[width=5cm]{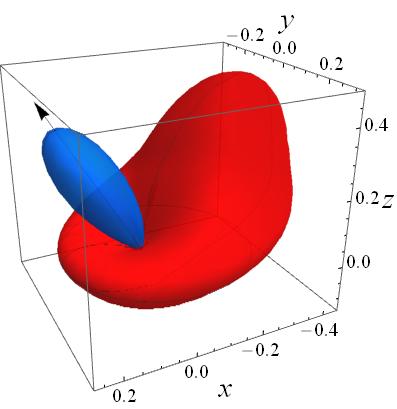} \\
\includegraphics[width=5cm]{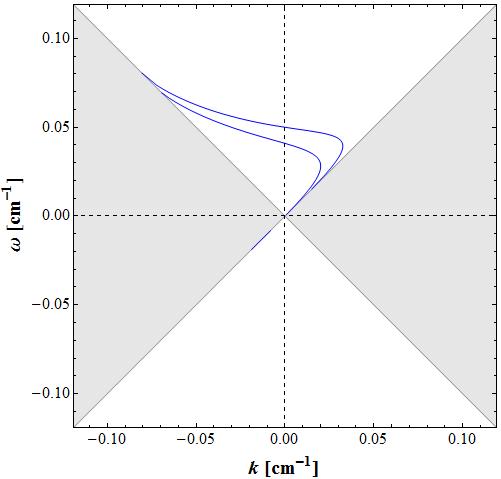} & \includegraphics[width=5cm]{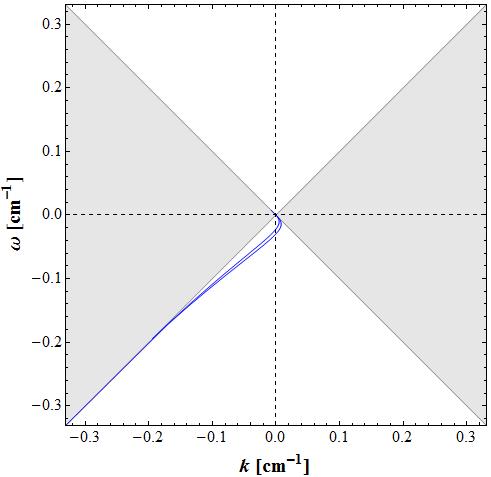} & \includegraphics[width=5cm]{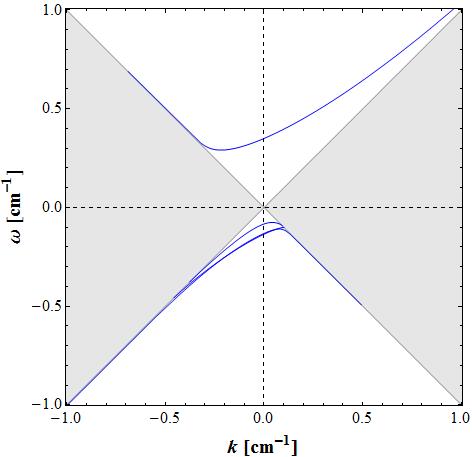}\\
\includegraphics[width=5cm]{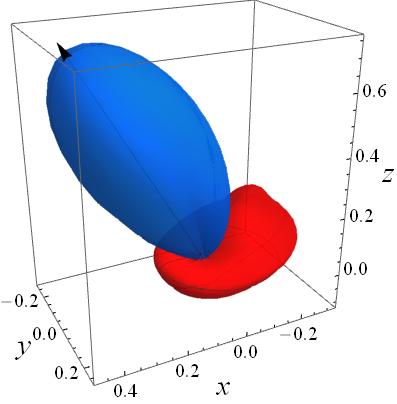} &\includegraphics[width=5cm]{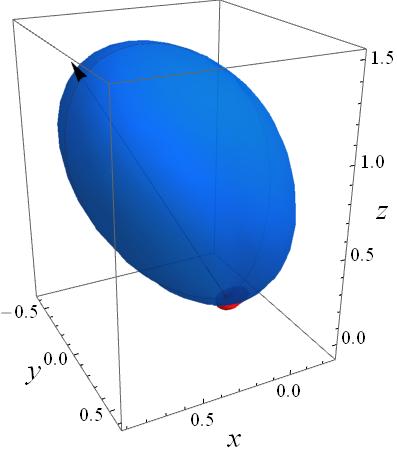} & \includegraphics[width=5cm]{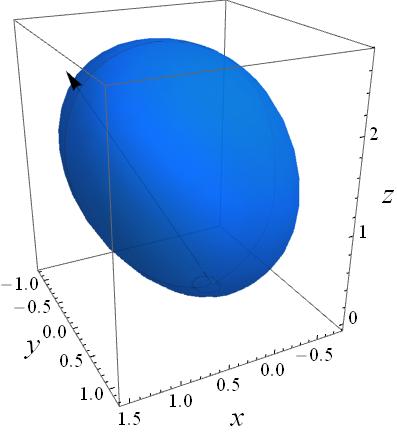}\\
 \end{tabular}\caption{\label{DR-Rend2759} Same as Figs.~\ref{DR-Rend15} and \ref{DR-Rend1904} for modified data at $t_{pb}$ = 275.9 ms. The multiplication factor is 1.2, 1.3 and 1.4 for the top two rows while it is 1.5, 1.7 and 2.0 for the bottom two rows. from left to right. The direction of $\mathbf{k}$ is set to the crossing direction, which are indicated by black arrows.} 
\end{figure*}

\begin{figure*}[t]
\begin{tabular}{ccc}
\includegraphics[width=5cm]{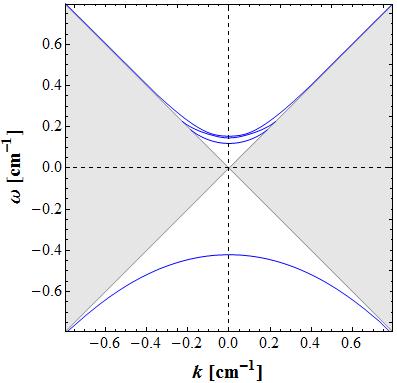} & \includegraphics[width=5cm]{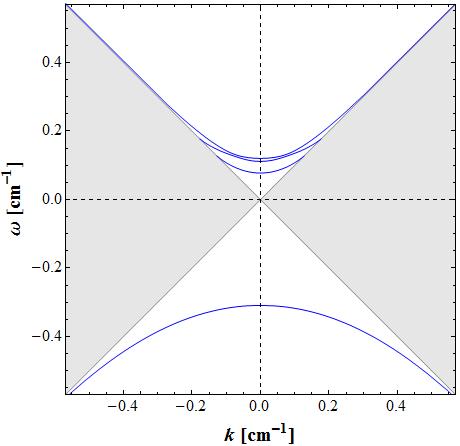} & \includegraphics[width=5cm]{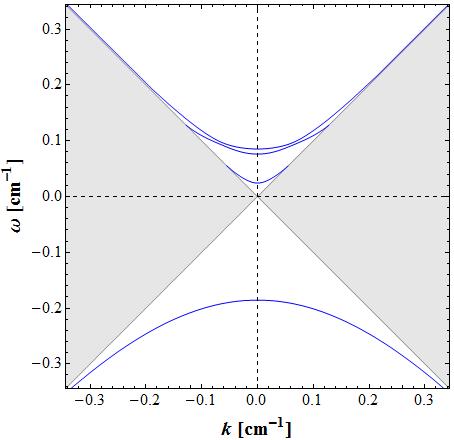}\\
\includegraphics[width=5cm]{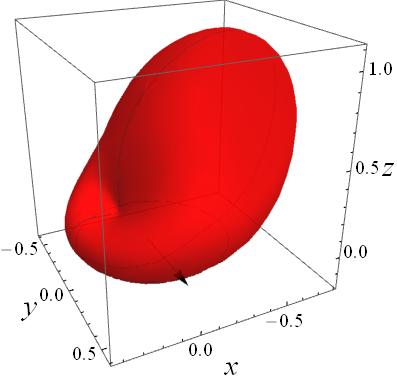} & \includegraphics[width=5cm]{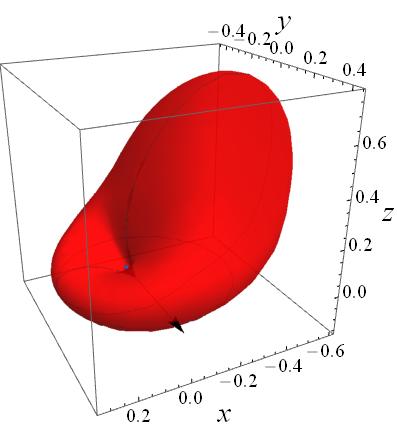} & \includegraphics[width=5cm]{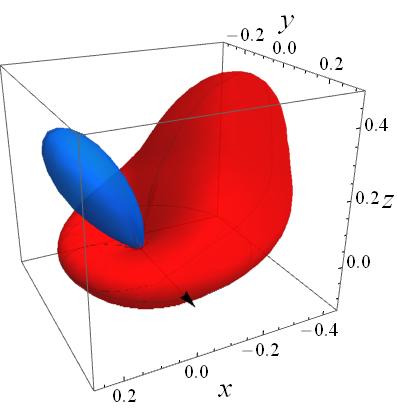} \\
\includegraphics[width=5cm]{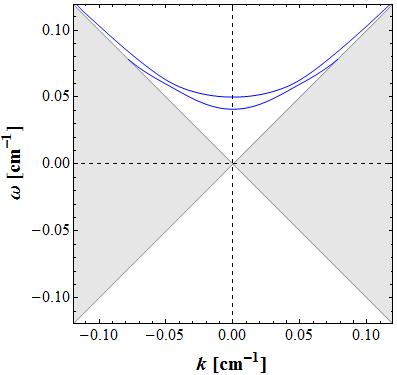} & \includegraphics[width=5cm]{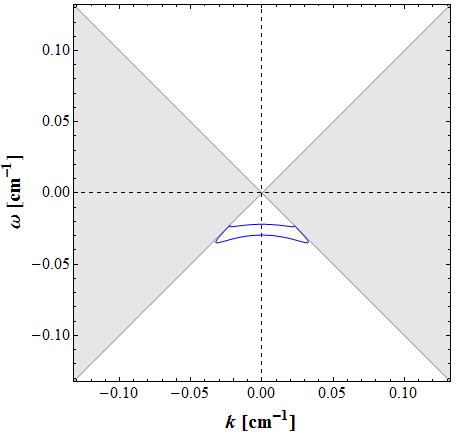} & \includegraphics[width=5cm]{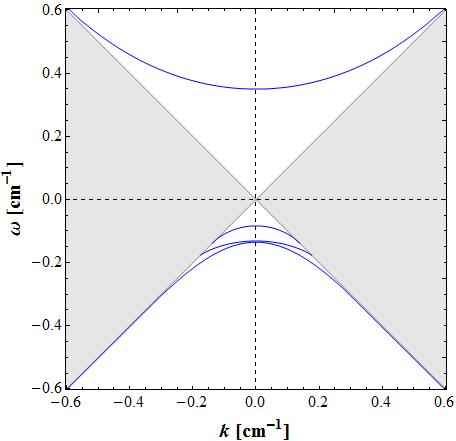}\\
\includegraphics[width=5cm]{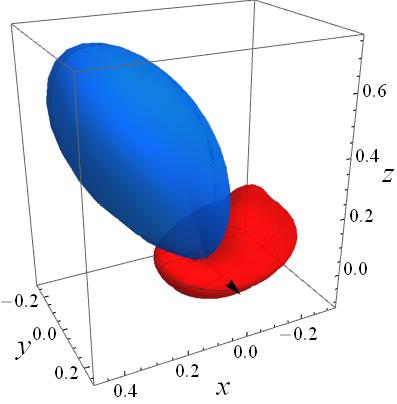} &\includegraphics[width=5cm]{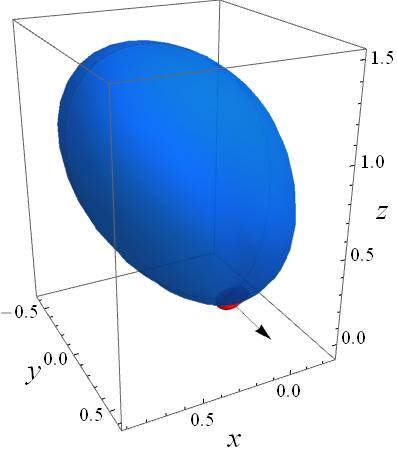} & \includegraphics[width=5cm]{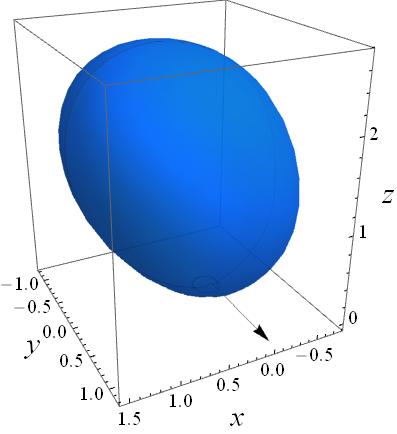}\\
 \end{tabular}\caption{\label{DR-Rend2759-y} Same as Fig.~\ref{DR-Rend2759} but for $\mathbf k$ in the y-direction as shown by black arrows. The multiplication factor is 1.2, 1.3 and 1.4 for the top two rows while it is 1.5, 1.7 and 2.0 for the bottom two rows from left to right.} 
\end{figure*}

\begin{figure*}[t]
\begin{tabular}{cc}
\includegraphics[width=7cm]{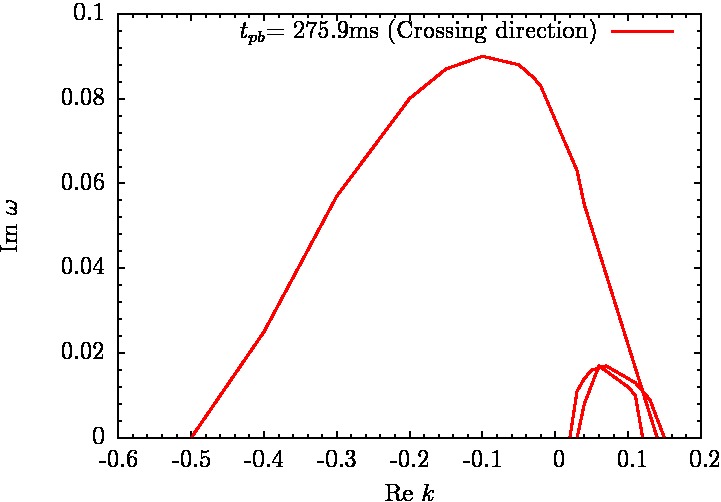} & \includegraphics[width=7cm]{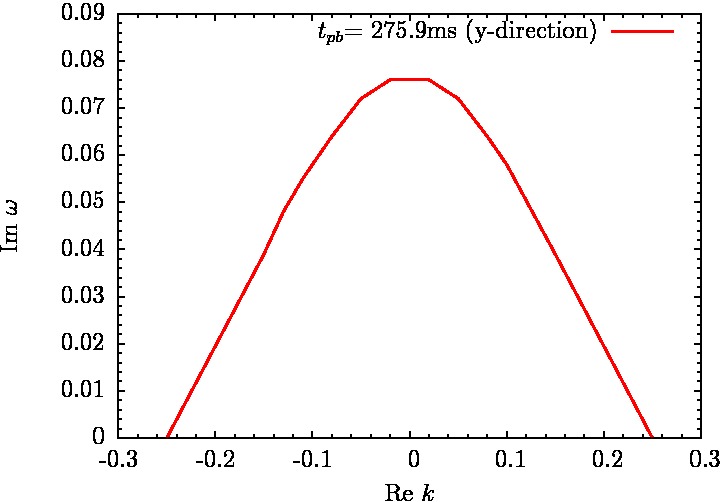}\\

 \end{tabular}\caption{\label{growth-2759} A comparison of the growth rates as a function of $k$ between the crossing and radial directions at $t_{pb}$ = 275.9 ms. The multiplication factor is 1.5, which corresponds to the near equal populations of $\nu_e$ and $\bar{\nu}_e$.}
\end{figure*}

As mentioned earlier, the DR is actually a function of the direction of $\mathbf k$. We hence study another direction, i.e., the local radial direction, which is the most common choice in the literature. We show in Fig.~\ref{DR-Rend15-radial} the DR's together with the angular distribution differences for this choice of the direction of $\mathbf k$. Note that the latters are the same as those in Fig.~\ref{DR-Rend15} except for the arrows that show the direction of $\mathbf k$. It is evident that the DR's do not change much, with a gap always open in $\omega$ and no peak in $k$, as we increase the multiplication factor. Interestingly, we still find instability after crossing. The growth rate is initially  much smaller than that in the crossing direction but increases with the multiplication factor and becomes comparable at the nearly equal populations of $\nu_e$ and $\bar\nu_e$, which is shown in Fig.~\ref{growth-new}. This is the reason why we chose the crossing direction for the linear analysis of the original \mbox{ numerical data earlier.} Although not presented, we have explored other, randomly chosen directions of $\mathbf{k}$ at the equal populations and observed that the instability occurs widely with similar growth rates. It is also found from the comparison of Fig.~\ref{growth-new} with Fig.~\ref{growth} that the maximum growth rate is greater at the equal populations than at the first crossing. As the multiplication factor gets even larger and $\bar\nu_e$ becomes dominant over $\nu_e$, the maximum growth rates for both the crossing and radial directions decreases again and drop to zero suddenly at the other crossing when the population of $\bar\nu_e$ overwhelms that of $\nu_e$ in all directions. We will demonstrate these features again later for another post-bounce time.

One thing should be mentioned here. We have so far defined the crossing direction to be the direction, in which the crossing occurs for the first time as the multiplication factor increases. This is fine until the crossing occurs but, after that, the crossing direction should be the direction, in which the distributions of $\nu_e$ and $\bar\nu_e$ are actually equal to each other and will vary with the multiplication factor in fact. If the growth rate is largest in the true crossing direction (this is remaining to be demonstrated), the maximum growth rates given in the previous paragraph may be smaller than the actual maximum growth rate. According to our small survey, however, this seems not a serious underestimation except in the close vicinity of the end of crossing. We will hence use the original definition of the crossing direction in the following even when $\bar\nu_e$ is dominant over $\nu_e$.

The motion of complex $k$ solutions is displayed in Fig.~\ref{newkplane}. This time we pick up Re $\omega$ = 0.1 $\mathrm{cm^{-1}}$ and vary the imaginary part of $\omega$ as Im $\omega$ = 0, 0.02 \mbox{and 0.07 $\mathrm{cm^{-1}}$} from left to right in the figure. As we can see, one of the solutions marked with red circles does cross the real axis but with a smaller value of Im $\omega$ compared with that for the crossing direction. Note that all lines rotate counter-clockwise this time because Re $\omega$ is positive. 

We extend our investigation to $t_{pb}$ = 190.4 ms. The first and second rows of Fig.~\ref{DR-Rend1904} display the DR's and angular distribution differences for the multiplication factor of 1.5, 1.6 and 1.7, respectively, among which the last one roughly corresponds to the occurence of the crossing for the first time in this case. Note that this factor is much smaller than the previous one, 35, for $t_{pb}$ = 15.0 ms, since the asymmetry in the population is much smaller between $\nu_e$ and $\bar\nu_e$ at this late time. It should be also mentioned that the fluxes of $\nu_e$ and $\bar\nu_e$ are much more aligned with each other as well as with the radial direction. In fact the crossing direction is $\theta_{\mathbf k}$ = 25.71~degree, which is chosen in drawing the DR in this case.

As we can see, the DR is changed qualititavely also in this case. In fact, it is quite similar to the change in the previous case. Nothing qualititavley different happens until the crossing occurs as shown in the first two pictures on the first row. The DR is changed rather aruptly near the multiplication of 1.7 for the third picture on the same row. In the left most picture on the third row, $\nu_e$ and $\bar\nu_e$ are abundant nearly equally but the DR is different from that in the corresponding case for $t_{pb}$~=~15.0~ms. (see the leftmost picture on the third row in Fig.~\ref{DR-Rend15}). In the second and third pictures on the third row, $\bar\nu_e$ dominates over $\nu_e$ in abundance, which is unlikely to occur in reality, and the DR returns to the one (but reflected with respect to Re $k$~=~0) in the top left panel. This is just as expected if one considers the symmetry between $\nu_e$ and $\bar\nu_e$.

At the multiplication factor 1.7, where the crossing occurs, we find the instability also in this case. The growth rates are displayed in the middle panel of Fig.~\ref{growth} as a function of $k$. There are again four branches corresponding to the peaks in $k$ observed in the DR. Shown in Fig.~\ref{DRmod1904} is the movements of complex $k$ solutions for $\omega$ = -0.75 $\mathrm{cm^{-1}}$ in the complex $k$-plane. As the value of the imaginary part of $\omega$ increases, two complex solutions, one above and the other below the real axis, move upwards and the latter reaches the real axis as shown in the middle and right panles of the same figure.

\begin{figure*}[t]
\begin{tabular}{ccc}
\includegraphics[width=4.85cm]{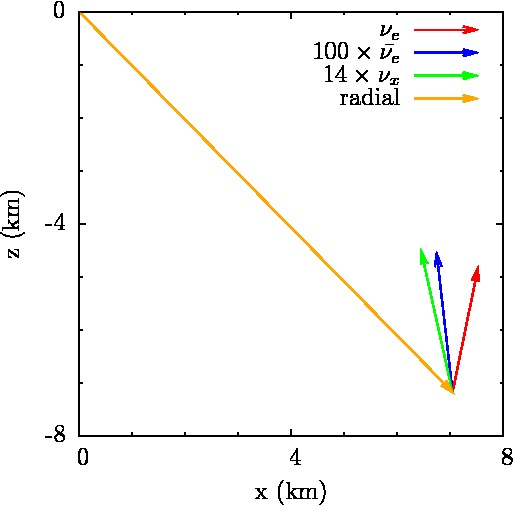} &\includegraphics[width=5cm]{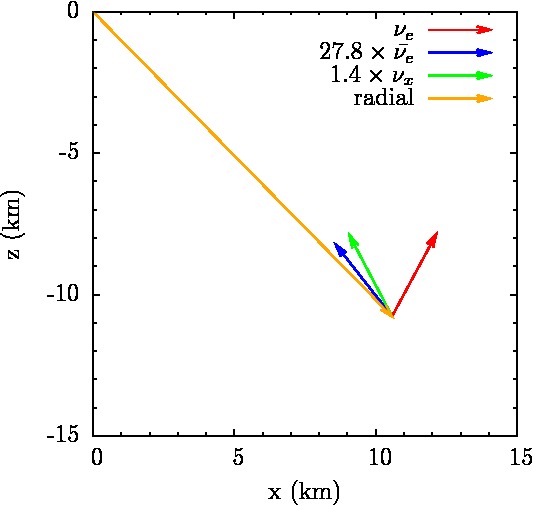} & \includegraphics[width=5cm]{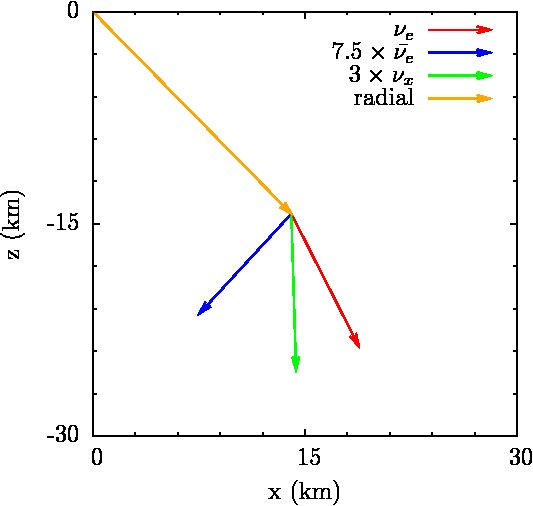}\\
\includegraphics[width=5cm]{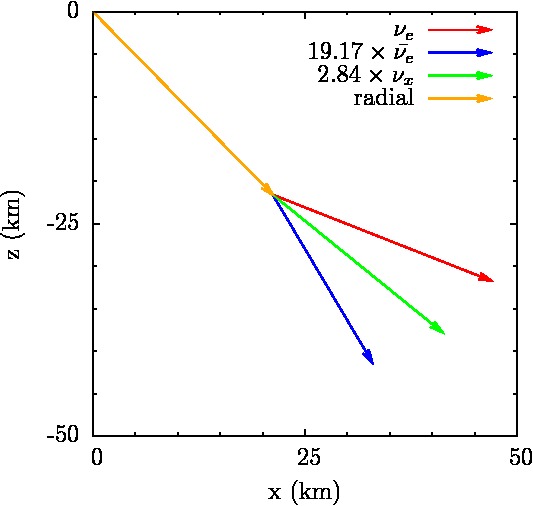} & \includegraphics[width=5cm]{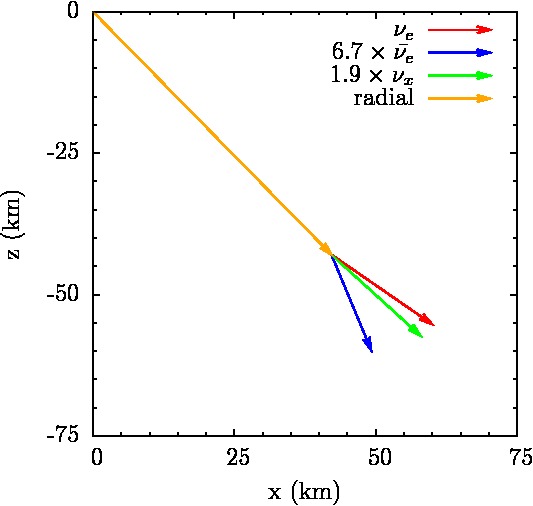} & \includegraphics[width=5cm]{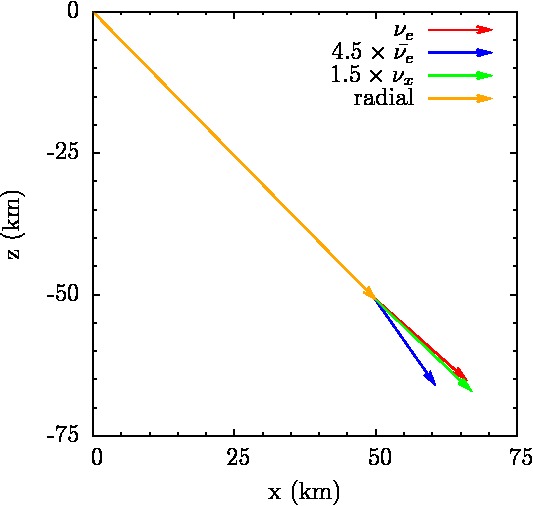}\\
\end{tabular}\caption{\label{flux15ms_30km.jpg} Neutrino fluxes of different species at different points both inside and outside the PNS at $t_{pb}=$ 15.0 ms. The spatial points are indicated by yellow arrows, which also show the local radial directions. Red, blue and green arrows display the fluxes of $\nu_e$, $\bar\nu_e$ and $\nu_x$, respectively. Note that they are normalized as shown in each panel.} 
\end{figure*}

Finally, we move on to $t_{pb}$ = 275.9 ms, which is quite similar to the $t_{pb}$ = 190.4 ms case as understood from Figs.~\ref{275snap}, \ref{flux2759} and \ref{DRreal15}. Figure~\ref{DR-Rend2759} gives the DR's and the angular distribution differences for the crossing direction, which is $\theta_{\mathbf k}$ = 30 degree in this case. At the crossing, which occurs slightly before the top right panel of this figure, the instability emerges. The growth rates are presented as a function of $k$ in the right panel of Fig.~\ref{growth}. There are four branches again, corresponding to the peaks in the DR. Since other features of the DR are not much different from those for $t_{pb}$ = 190.4 ms, here we look at another direction of $\mathbf k$, which we choose to be the positive \mbox{y-direction.}

We show the DR's and the angular distribution differences for this case in Fig.~\ref{DR-Rend2759-y}. As in the case for the \mbox{radial $\mathbf{k}$} at $t_{pb}$ = 15.0 ms, the DR does not change much with the multiplication factor. As a matter of fact, it looks essentially unchaged at the first crossing (see the top right panel). As a result, no unstable solution is found in this direction at this point. We still find instability, however, for a bit larger multiplication factor as shown in Fig.~\ref{growth-2759}, in which the growth rates are shown as a function of $k$ for this direction at the multiplication factor of 1.5. i.e., when $\nu_e$ and $\bar\nu_e$ are populated roughly equally (see the bottom left panel). For comparison we include the results for the crossing direction for the same multiplication factor. It is again found that the instability grows faster in the crossing direction and that the maximum growth rate is greater at the equal population than at the first crossing. Interestingly the instability still exists and the growth rates are even higher at the  multiplication factor of 1.7 (the middle panels on the third and fourth rows), where $\bar\nu_e$ evidently dominates $\nu_e$ very much. Then the growth rate drops to zero rather suddenly when the population of $\bar\nu_e$ is larger than that of $\nu_e$ in all directions. These results demonstrate again that a mere inspection of DR is not sufficient to detect instability and suggest that the growth rate depends on the direction of $\mathbf k$ and the angular distributions of $\nu_e$ and $\bar\nu_e$ in a subtle way. We certainly need more systematic investigation, though, to see how generic this result is.

\begin{figure*}[t]
\begin{tabular}{cc}
\includegraphics[width=5cm]{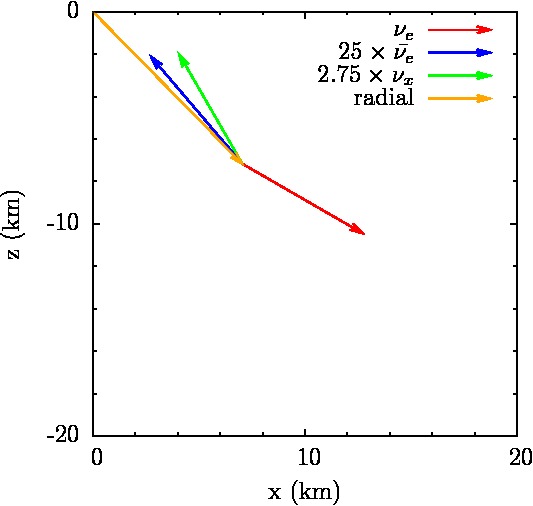} &\includegraphics[width=5cm]{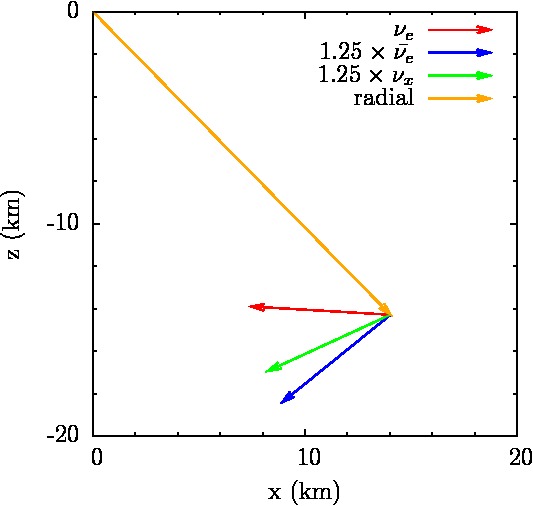}\\

\includegraphics[width=5cm]{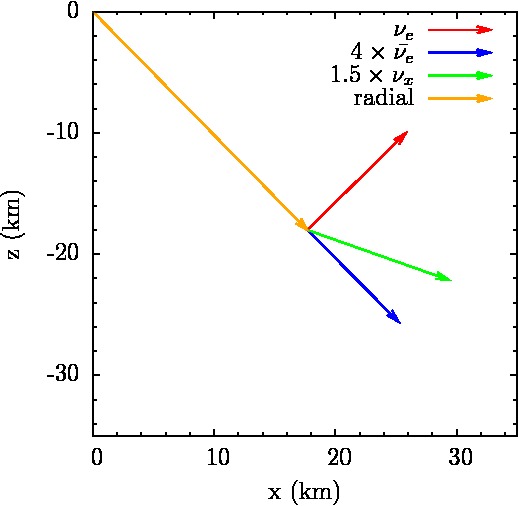} &\includegraphics[width=5cm]{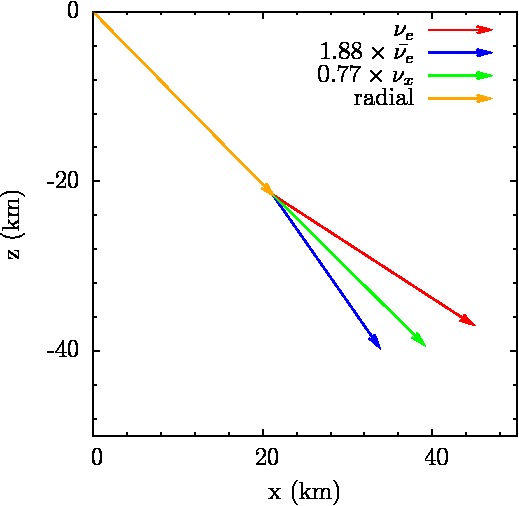}\\
\end{tabular}\caption{\label{flux1904ms_30km.jpg} Same as Fig.~\ref{flux15ms_30km.jpg} but at $t_{pb}=$ 190.4 ms.} 
\end{figure*}

\begin{figure*}[t]
\begin{tabular}{cc}
\includegraphics[width=5cm]{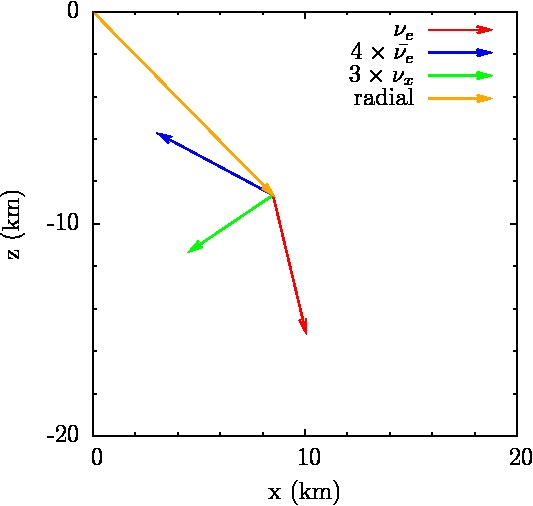} &\includegraphics[width=5cm]{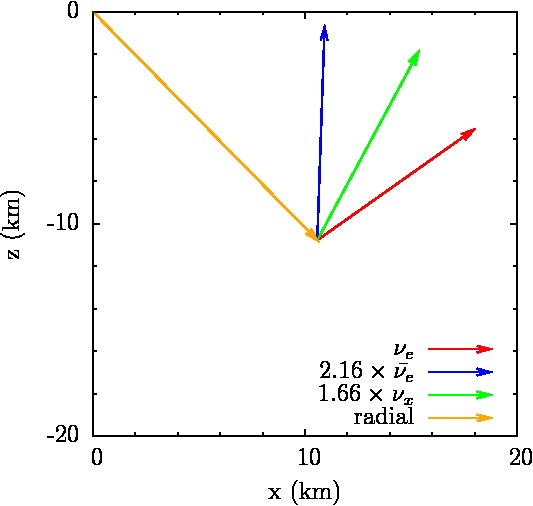}\\

\includegraphics[width=5cm]{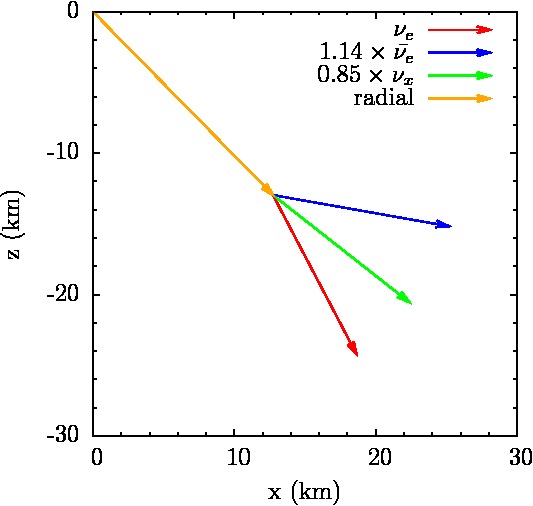} &\includegraphics[width=5cm]{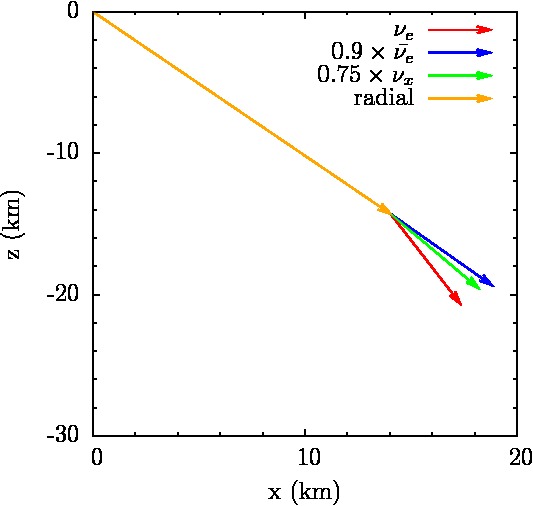}\\
\end{tabular}\caption{\label{flux2759_25km.jpg} Same as Figs.~\ref{flux15ms_30km.jpg} and \ref{flux1904ms_30km.jpg} but at $t_{pb}=$ 275.9 ms.} 
\end{figure*}

\section{Summary and Discussions}

In this paper, we have conducted a pilot study on the possibilities of the so-called fast-pairwise \mbox{collective} neutrino oscillations in the supernova core, applyting the linear stability analysis to a few data selectively extracted from the fully self-consistent realistic \mbox{Boltzmann-neutrino-radiation-hydrodynamical simulation} for the non-rotating progenitor of 11.2$M_{\odot}$. We have obtained the neutrino distributions at the point near the neutrino sphere ($r$ = 44.8 km, $\theta$ = 2.36 radian) from the numerical data at three different post-bounce times: $t_{pb}$ = 15, 190.4 and 275.9 ms. This is in fact one of the places, where we found the largest misalignment between the fluxes of $\nu_e$ and $\bar\nu_e$ at \mbox{$t_{pb}$ = 15.0 ms}. This misalignment is produced mainly by convective motions and is treated consistently with hydrodynamics by our \mbox{Boltzmann-neutrino-radiation-hydrodynamics} code. At the later times, the misalignment is much reduced as the neutrino sphere retreats to smaller radii. We suppose that this is effictively similar to going to larger radii, where the neutrino distributions become forward-peaked and get more or less aligned with each other and with the radial direction. We did not go deeper inside the neutrino sphere, since in the linear analysis of the neutrino oscillations we ignored all neutrino interactions other than the forward scatterings that induce the refractive effect. We have conducted linear analysis then, solving the equation for the dispersion relation to obtain complex solutions, which are supposed to indicate the instability that instigates the flavor conversion. 

It turns out that we have found no unstable mode in any case. The dispersion relations are qualitatively the same in the three cases we studied: there are a few branches in general and a gap is open in $\omega$, the frequency of the perturbation. This is an indication of the existence of complex~$k$ (the wave number of the perturbation) \mbox{solutions} for some real values of $\omega$, which we have indeed confirmed. We are interested, however, in the complex $\omega$ solution for real $k$ rather than the complex $k$ solution for real $\omega$. In none of the three cases we have found such a solution.  

In order to elucidate what was lacking, we have modified the distributions of $\bar\nu_e$, multiplying the arbitrary factors with the original distribution functions. We have repeated the same analysis for these modified data and demonstrated that the unstable modes start to exist once the crossing occurs in the angular distributions of $\nu_e$ and $\bar\nu_e$ for all three cases. This seems to confirm the conventional expectation that such crossing is the condition for the fast-pairwise conversion. Note, however, that this was not demonstrated so far in non-spherical settings, in which neutrino fluxes are misaligned with each other. It should be also stressed that the DR depends on the direction of $\mathbf k$, the wave vector of perturbation. In these analyses we have chosen the crossing direction, i.e. the direction, in which the crossing is most likely. Our pilot study of other directions indicates indeed that this is likely to be the direction with the greatest \mbox{growth rate of instability.} 

We have explored in detail the DR's for these various cases and found that it changes qualitatively and rather suddenly at the point, where crossing happens, for $\mathbf{k}$ in the crossing direction. The gap in $\omega$ is closed and there appear some peaks in $k$ instead; the unstable modes are associated with these peaks. It is intriguing however, that such a change in the DR has not been observed for other directions of $\mathbf k$ such as the radial direction: the gap is still open in $\omega$ and there is no peak in $k$. We have still found instability in such cases. Note that the DR's before the crossing never have instability although they look very similar to that at the crossing in this case. This implies that a mere inspection of DR may not be sufficient to judge the existence of the unstable mode. It is added that the growth rates of the instabilities for other directions tend to be smaller than those for the crossing direction particularly near the threshold. For the nearly equal populations of $\nu_e$ and $\bar\nu_e$, on the other hand, the instability seems to occur in a wide direction with similar growth rates. Although our results seem to suggest that the crossing is the right criterion for the fast-pairwise conversion, it should be consolidated by more systematic investigations, possibly with simplified models (Morinaga et al. in preparation). In fact, we have seen different types of DR's in this paper but have not understood how they are related with the angular distributions of neutrinos and the direction of $\mathbf k$.  

This paper is meant to be a pilot study for more thorough and systematic explorations of the possibility of the fast-pairwise conversion in some region(s)of the core in some phase(s) of the entire supernova evolution. Although we have to wait for the detailed survey to get a firm conclusion, our pilot investigation indicates that it may not be so easy to obtain a crossing in the angular distributions of $\nu_e$ and $\bar{\nu}_e$. In fact, they can have quite different angular distributions near the neutrino sphere, where they are strongly coupled with matter in convective motions. Because of the high density, however, the population of $\bar\nu_e$ is strongly suppressed as is clear in our case for \mbox{$t_{pb}$ = 15.0 ms} (but see also \cite{Abbar:2018shq}). Then the crossing is impossible even if the fluxes are highly misaligned, since $\nu_e$ is dominant over $\bar\nu_e$ for all propagation directions. As the radius increases and the density decreases, $\bar\nu_e$ gets more populated but the fluxes for $\nu_e$ and $\bar\nu_e$ become more aligned with each other at the same time. Then the crossing is difficult again as we saw in the cases for $t_{pb}$ = 190.4 and \mbox{275.9 ms}. In order to demonstrate these situations, we show the energy-integrated fluxes of three neutrino species at other points, both inside and outside the neutrino sphere, for the same three post-bounce times in Figs.~\ref{flux15ms_30km.jpg}, \ref{flux1904ms_30km.jpg} and \ref{flux2759_25km.jpg}. As explained above, either the large misalignment is accompanied by the great asymmetry in the population or the alignment occurs inevitably with the realization of almost equal populations. 

What all these examples show is the fact that the crossing is a highly subtle thing \cite{Abbar:2018shq}. We are currently undertaking this project, employing different numerical data obtained in our Boltzmann simulations for other progenitor models and nuclear equations of state. The effects of stellar rotation and PNS kick are also being investigated. It is stressed again that our simulations are fully \mbox{self-consistent} in the sense that neutrino transfer is computed not as a post-process as in the preceding work \cite{Abbar:2018shq} but simultaneously with hydrodynamics. Since the neutrino distributions are the single most important ingredient for the fast-pairwise conversion, we believe that such consistency to treat neutrino transport and hydrodynamics is crucial. Although we adopted the 2-flavor approximation in this paper, the extension of the formulation to 3-flavor is certainly necessary. The possibility of the \mbox{fast-pairwise} collective neutrino oscillations in our latest three dimensional model computed with our Boltzmann code is also under study and the results will be reported soon \mbox{(Delfan Azari et al. in preparation)}. Last but not least, the criterion for the instability we employed in this paper may be imperfect and mathematically more rigorous treatment is in order \cite{2017PhRvD..96d3016C}. We have made an interesting progress also in this respect and will present it soon (Morinaga et al. in preparation).   

\begin{acknowledgments}

M.D.A was supported by the Ministry of \mbox{Education}, Culture, Sports, Science and Technology of Japan (MEXT) and Waseda University for his \mbox{postgraduate studies.} \mbox{H.N. was} supported by Princeton University through DOE SciDAC4 Grant \mbox{DE-SC0018297} \mbox{(subaward 00009650).} This work is also supported by HPCI Strategic Program of Japanese MEXT and K computer at the RIKEN (Project ID: hpci 160071, 160211, 170230, 170031, 170304, hp180179, hp180111). This work is supported by Grant-in-Aid for Scientific Research (26104006, 15K05093)
and Grant-in-Aid for Scientific Research on Innovative areas "Gravitational wave physics and astronomy:Genesis" (17H06357, 17H06365) from the Ministry of Education, Culture, Sports, Science and Technology (MEXT), Japan. For providing high performance computing resources, Computing Research Center, KEK, JLDG on SINET4 of NII, Research Center for Nuclear Physics, Osaka University, Yukawa Institute for Theoretical Physics, Kyoto University, Nagoya University, and Information Technology Center, University of Tokyo are acknowledged. This work was partly supported by research programs at K-computer of the RIKEN AICS (Project ID: hp130025, hp140211, hp150225), HPCI Strategic Program of Japanese MEXT, “Priority Issue on Post-K computer” (Elucidation of the Fundamental Laws and Evolution of the Universe) and Joint Institute for Computational Fundamental Sciences (JICFus). The numerical computations were performed on the K computer, at AICS, FX10 at the Information Technology Center of Tokyo University, on SR16000 and Blue Gene/Q at KEK under the support of its Large Scale Simulation Program (14/15-17, 15/16-08, 16/17-11), on SR16000 at Yukawa Institute for Theoretical Physics, Kyoto University, Research Center for Nuclear Physics (RCNP) at Osaka University, and on the XC30 and the general common use computer system at the Center for Computational Astrophysics, CfCA, the National Astronomical Observatory of Japan. Large-scale storage of numerical data is supported by JLDG constructed over SINET4 of NII. 

\end {acknowledgments}

%\appendix
\appendix*
\section{}

\begin{figure}[t]
\begin{tabular}{c}
\includegraphics[width=5.5cm]{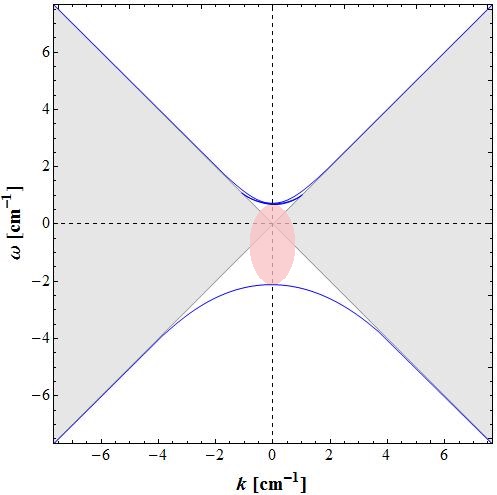}\\
\includegraphics[width=5.5cm]{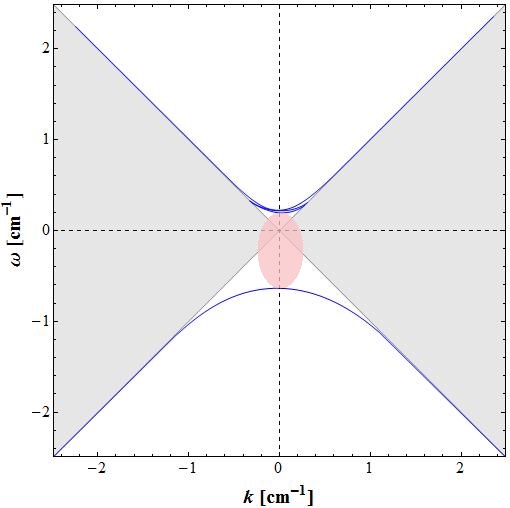}
\end{tabular}  \caption{\label{DRreal-later} Same as Fig.~\ref{DRreal15} but for different post-bounce times. Top panel corresponds to the ${t_{pb}}$~=~190.4 whereas the bottom panel corresponds to \mbox{${t_{pb}}$~= 275.9 ms}. Note that the scales are different from panel to panel.}
\end{figure}

\begin{figure*}[t]
\begin{tabular}{ccc}

\includegraphics[width=6cm]{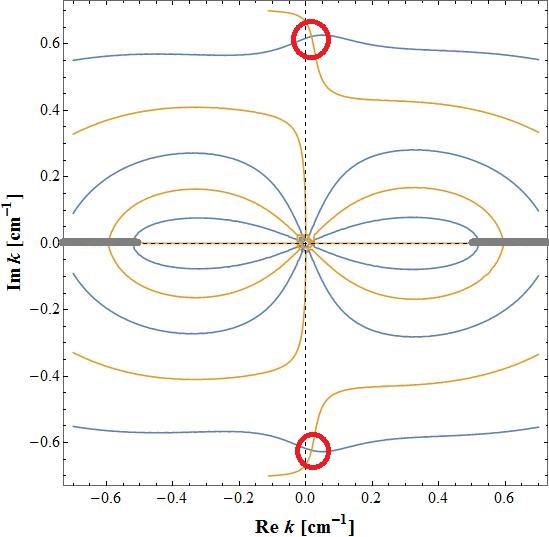} & \includegraphics[width=6cm]{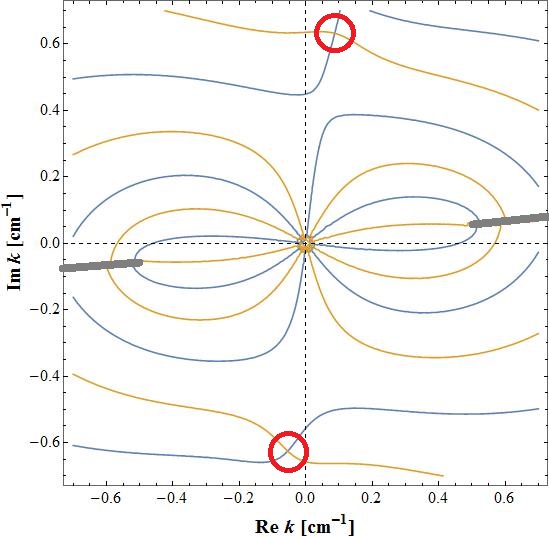} & \includegraphics[width=6cm]{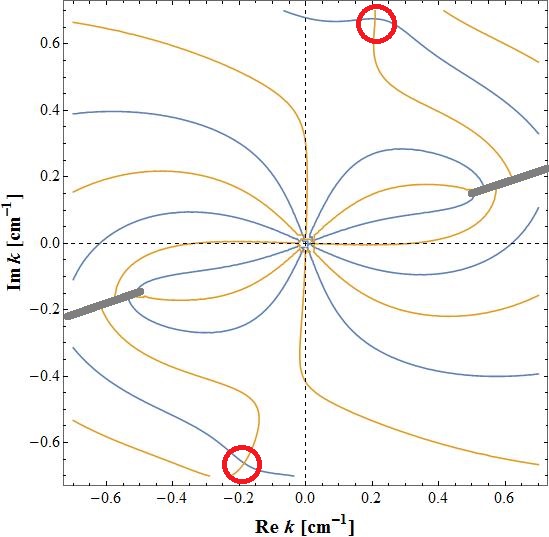}\\
\includegraphics[width=6cm]{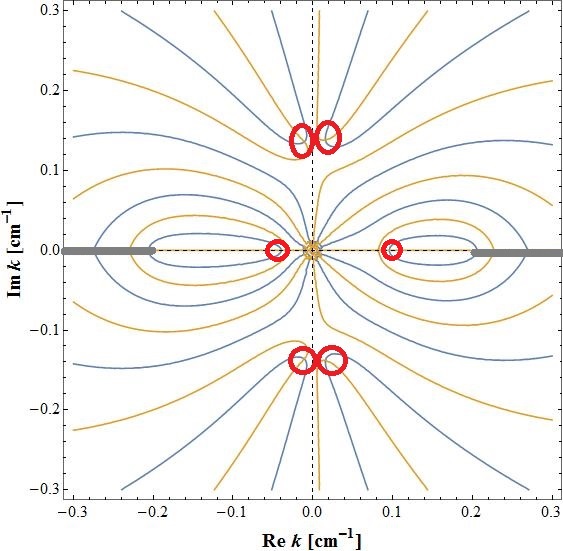} & \includegraphics[width=6cm]{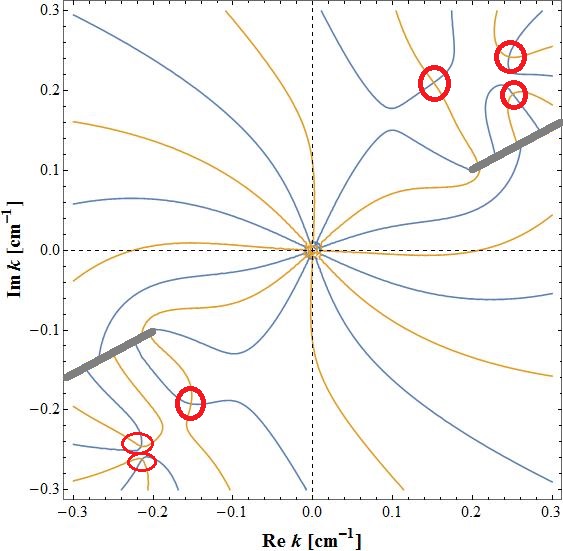} & \includegraphics[width=6cm]{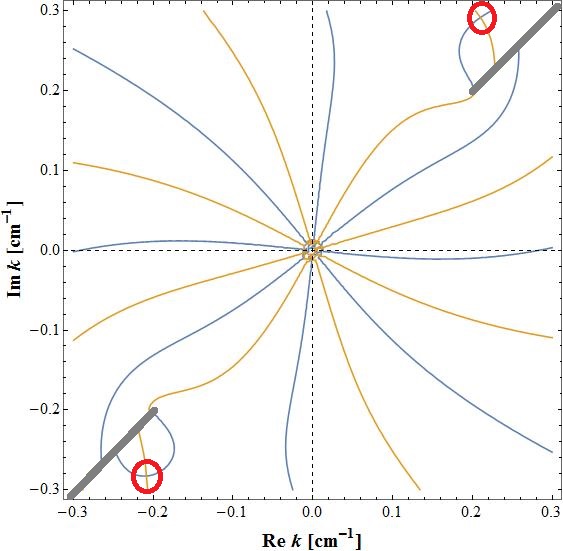}\\
\end{tabular}\caption{\label{DRroot} Same as Fig.~\ref{DRroot15} but at $t_{pb}$ = 190.4 ms (top panels) and for $t_{pb}$ = 275.9 ms (bottom panels). In the former case, we take Re $\omega=$ 0.5 $\mathrm{cm^{-1}}$ and Im $\omega=$ 0, 0.05 and 0.15 $\mathrm{cm^{-1}}$ from left to right whereas in the latter we set Re $\omega=$ 0.2 $\mathrm{cm^{-1}}$ and \mbox{Im $\omega=$ 0, 0.1 and 0.2 $\mathrm{cm^{-1}}$} from left to right.} 
\end{figure*}

We summarize the results for our analysis of the original data for $t_{pb}$ = 190.4 and 275.9 ms. The top panel of Fig.~\ref{DRreal-later} shows the DR for $t_{pb}$ = 190.4 ms, which is quite similar to the one for $t_{pb} = 15.0$ ms given in Fig.~\ref{DRreal15} except for the scale. We again pick up three different points, one inside the gap of the DR and the other two outside. We first choose a real value of $\omega$ = 0.5 $\mathrm{cm^{-1}}$, which is inside the DR gap (red region). The solutions of Re [det[$\Pi$]] = 0 and Im [det[$\Pi$]] = 0 in the complex $k$-plane are shown as blue and orange lines, respectively, in the top left panel of Fig.~\ref{DRroot}. Except for the scale, they are quite similiar to Fig.~\ref{DRroot15} for $t_{pb}$ = 15.0 ms just as expected. There are two complex solutions in $k$ near the imaginary axis. Also shown in the middle and right panels on the top row of Fig.~\ref{DRroot} are the same results but for Im $\omega$ = 0.05 and 0.15 $\mathrm{cm^{-1}}$, respectively.  As we can see again, there is no complex $k$ root that approaches the real $k$-axis at this time step, either. Other two points outside the DR gap, \mbox{$\omega$ = -4 and 2 $\mathrm{cm^{-1}}$}, are also investiagated. Just as in the case for $t_{pb}$ = 15.0 ms we do not obtain even a solution, not to mention a crossing, which is not shown here to avoid an unnessacry repetition.

Although the results are anticipated, we also apply the same analysis to the last time step, $t_{pb}$ = 275.9 ms, and check the  possibilities of the instability. The bottom panel of Fig.~\ref{DRreal-later} is the corresponding DR, which is similar to the previous ones. We pick up three different points, one inside the gap of the DR and the other two outside it. We choose $\omega$ = 0.2 $\mathrm{cm^{-1}}$ this time inside the DR gap. The left panel on the second row of Fig.~\ref{DRroot} shows the solutions in the complex $k$-plane for this real $\omega$. We exhibit also the results for Im $\omega$ = 0.1 and 0.2~$\mathrm{cm^{-1}}$ with the same Re $\omega$ in the middle and right panels on the same row. We find no crossings of the real $k$-axis at this time step, either. The other two points outside the DR gap, Re $\omega$ = -2 and 1~$\mathrm{cm^{-1}}$, do not have even a solution.

%\end {appendix*}
\bibliography{Papers_vf}

\end{document}